\documentclass[manuscript, acmsmall, screen]{acmart}
\AtBeginDocument{%
  }
    
\usepackage{xspace}
\usepackage{amsmath}
\usepackage{tcolorbox}
\usepackage{listings}
\usepackage{xcolor}
\usepackage{fancyvrb}
\usepackage{tabularx}
\usepackage{caption}
\usepackage{float}
\usepackage{graphicx}
\usepackage{pifont}
\usepackage{multirow}
\usepackage{colortbl}
\usepackage{hyperref}
\usepackage{booktabs, multirow} % for borders and merged ranges
\usepackage{soul}% for underlines
\usepackage{xcolor,colortbl} % for cell colors
\usepackage{changepage,threeparttable} % for wide tables
\usepackage[utf8]{inputenc}

\definecolor{deepgreen}{rgb}{0.0, 0.5, 0.0}
\definecolor{ForestGreen}{rgb}{0.0, 0.5, 0.0}

\newcommand{\phead}[1]{\noindent {\bf #1}}
\usepackage{enumitem}

\newcommand{\raw}{{\textit{Function-Only}}\xspace}
\newcommand{\tool}{\textit{RobuNFR}\xspace}
\newcommand{\first}{{\textit{NFR-Integrated}}\xspace}

\newcommand{\functional}{{\textit{Functional}}\xspace}
\newcommand{\second}{{\textit{NFR-Enhanced}}\xspace}

\newcommand{\claudeNew}{Claude3.5-1022\xspace}
\newcommand{\claudeOld}{Claude3.5-0620\xspace}
\newcommand{\gptThreeOld}{GPT3.5-1106\xspace}
\newcommand{\gptThreeNew}{GPT3.5-0125\xspace}
\newcommand{\gptFourOld}{GPT4o-0513\xspace}
\newcommand{\gptFourNew}{GPT4o-0806\xspace}

\newcommand\greybox[1]{%
  \vskip\baselineskip%
  \vspace{-0.8\baselineskip}
  \par\noindent\colorbox{lightgray}{%
    \begin{minipage}{\linewidth}#1\end{minipage}%
  }%
  \vspace{-0.8\baselineskip}
  \vskip\baselineskip%
}

\newcommand{\rqboxc}[1]{\begin{tcolorbox}[left=4pt,right=4pt,top=4pt,bottom=4pt,colback=gray!35,colframe=gray!35,before skip=3pt,after skip=3pt]#1\end{tcolorbox}}

\definecolor{codegreen}{rgb}{0,0.6,0}
\definecolor{codegray}{rgb}{0.5,0.5,0.5}
\definecolor{codepurple}{rgb}{0.58,0,0.82}
\definecolor{backcolour}{rgb}{0.95,0.95,0.92}
\definecolor{lightgray}{gray}{0.9}
\definecolor{lightergray}{gray}{0.95}

\lstdefinestyle{mystyle}{
    backgroundcolor=\color{backcolour},   
    commentstyle=\color{codegreen},
    keywordstyle=\color{magenta},
    numberstyle=\tiny\color{codegray},
    stringstyle=\color{codepurple},
    basicstyle=\ttfamily\footnotesize,
    breakatwhitespace=false,         
    breaklines=true,                 
    captionpos=b,                    
    keepspaces=true,                 
    numbers=left,                    
    numbersep=5pt,                  
    showspaces=false,                
    showstringspaces=false,
    showtabs=false,                  
    tabsize=2
}

\acmJournal{TOSEM}
\acmVolume{XX}
\acmNumber{X}
\acmYear{2025}
\setcopyright{acmlicensed}
\copyrightyear{2025}
\acmDOI{XXXXXXX.XXXXXXX}
\acmISBN{978-1-4503-XXXX-X/18/06}

\begin{document}

\title{\tool: Evaluating the Robustness of Large Language Models on Non-Functional Requirements Aware Code Generation}

\author{Feng Lin}
\orcid{0009-0009-3887-7071}
\affiliation{%
  \institution{Software PErformance, Analysis, and Reliability (SPEAR) lab, Concordia University}
  \city{Montreal}
  \country{Canada}
}
\email{feng.lin@mail.concordia.ca}

\author{Dong Jae Kim}
\orcid{0000-0002-3181-0001}
\affiliation{%
  \institution{DePaul University}
  \city{Chicago}
  \country{USA}
}
\email{dkim121@depaul.edu}

\author{Zhenhao Li}
\orcid{0000-0002-4909-1535}
\affiliation{%
  \institution{York University}
  \city{Toronto}
  \country{Canada}
}
\email{lzhenhao@yorku.ca}

\author{Jinqiu Yang}
\orcid{0000-0003-4282-406X}
\affiliation{%
  \institution{Concordia University}
  \city{Montreal}
  \country{Canada}
}
\email{jinqiu.yang@concordia.ca}

\author{Tse-Hsun (Peter) Chen}
\orcid{0000-0003-4027-0905}
\affiliation{%
  \institution{Software PErformance, Analysis, and Reliability (SPEAR) lab, Concordia University}
  \city{Montreal}
  \country{Canada}
}
\email{peterc@encs.concordia.ca}

\renewcommand{\shortauthors}{Feng et al.}

\begin{abstract}
When using LLMs to address Non-Functional Requirements (NFRs), developers may behave differently (e.g., expressing the same NFR in different words). Robust LLMs should output consistent results across these variations; however, this aspect remains underexplored. We propose \tool for evaluating the robustness of LLMs in NFR-aware code generation across four NFR dimensions—design, readability, reliability, and performance—using three methodologies: prompt variation, regression testing, and diverse workflows. Our experiments show that \tool reveals robustness issues in the tested LLMs when considering NFRs in code generation. Specifically, under prompt variation, including NFRs leads to a decrease in Pass@1 by up to 39\% and an increase in the standard deviation from 0.48 to 2.48 compared to the baseline without NFRs (i.e., Function-Only). While incorporating NFRs generally improves overall NFR metrics, it also results in higher prompt sensitivity. In regression settings, some LLMs exhibit differences across versions, with improvements in one aspect (e.g., reduced code smells) often accompanied by regressions in another (e.g., decreased correctness), revealing inconsistencies that challenge their robustness. When varying workflows, the tested LLMs show significantly different NFR-aware code generation capabilities between two workflows: (1) integrating NFRs and functional requirements into the initial prompt and (2) enhancing Function-Only-generated code with the same NFR.
\end{abstract}

\maketitle

\section{Introduction}
Code quality is a critical aspect of software development, alongside code correctness, and requires developers to consider Non-Functional Requirements (NFRs) throughout the development process. With the rise of Large Language Models (LLMs)~\cite{zhang2024surveylargelanguagemodels}, many LLM-based services, such as ChatGPT~\cite{chatGPT}, GitHub Copilot~\cite{copilotwebsite}, and Cursor~\cite{cursor}, have become increasingly integrated into modern development workflows. These services are now widely adopted for addressing code generation tasks related to NFRs, either by generating new code with NFR considerations in mind or by enhancing existing code to meet those requirements.
% https://docs.google.com/presentation/d/1s5D7iowsujq7Mr5XTLVv3j3WY8FbARtrGnIMFsudjQU/edit#slide=id.g31e2dfd0c01_0_1
\begin{figure*}
 \centering
\includegraphics[width=0.95\linewidth]{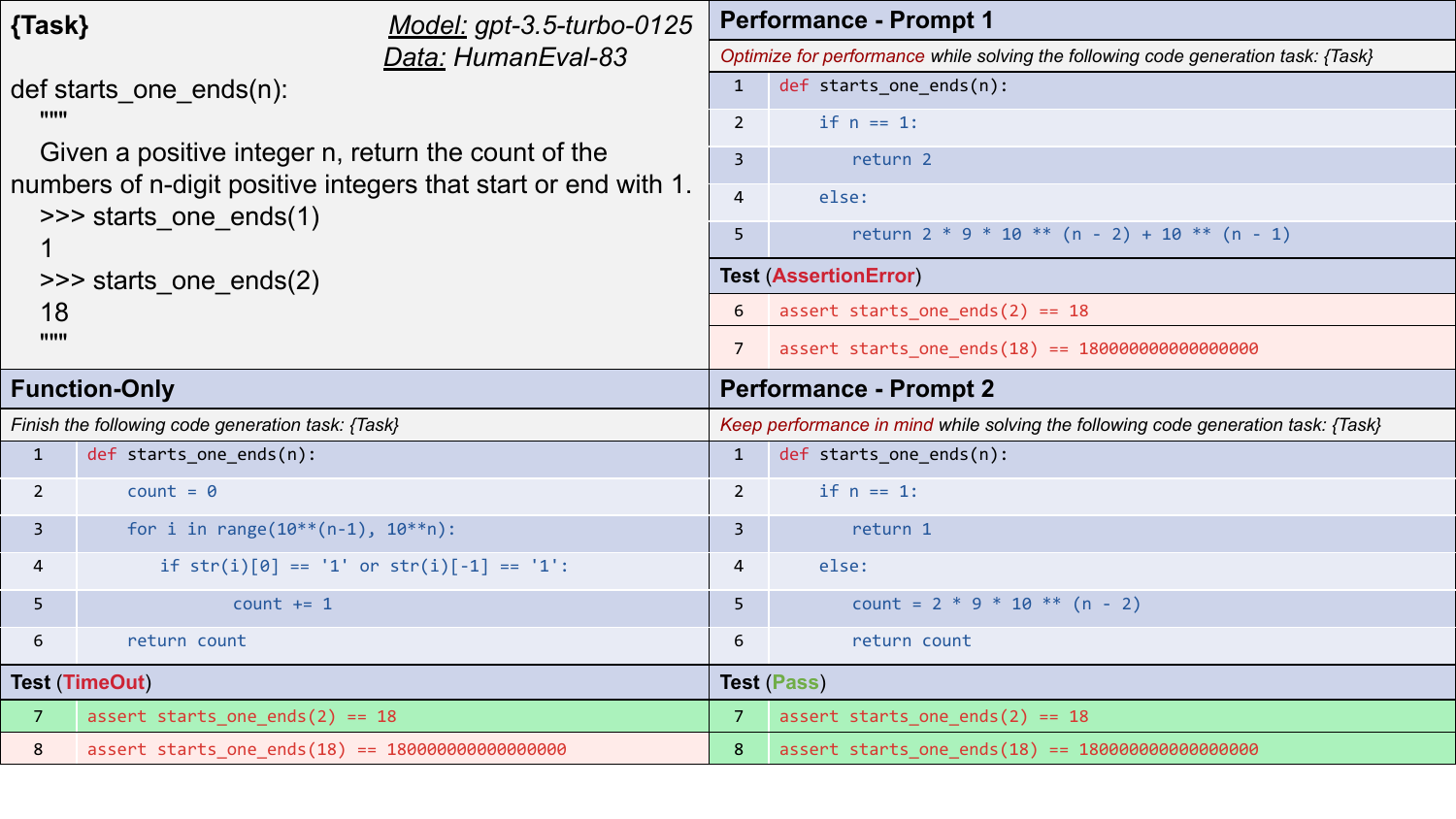}
 \caption{Simplified examples of Generated Code: One Without Performance Considerations, and Two With Performance Considerations Using Different Prompts.}
 \label{fig:intro_example}
 \vspace{-8mm}
 \end{figure*}

% \peter{do they just consider perf or do they consider other aspects?}Previous studies have examined both the functional correctness and overall quality of generated code. They assess whether the code not only passes all provided test cases but also achieves improved execution speed~\cite{singhal2024nofunevalfunnycodelms, waghjale2024eccoimprovemodelgeneratedcode}. 
% However, 
% \peter{can remove the sentences before}
In real-world scenarios, users employ LLMs to address NFRs in various ways. For example, different users might use distinct prompts to express the same NFRs, and these variations can lead to different outcomes, depending on the robustness of the LLM’s NFR-aware code generation capabilities. 
Figure~\ref{fig:intro_example} illustrates this scenario: for the same coding problem, the outcomes vary depending on whether performance is considered, and further differences emerge when different prompts are used to address performance requirements. Even when generated by the same LLM, code generated without explicit performance requirements relies on exhaustive iteration, leading to a \textit{Timeout} for large loop sizes. Moreover, even when performance is considered, \textit{Prompt 1} misleads the LLM into generating code that improves speed but raises an \textit{AssertionError}, whereas \textit{Prompt 2} achieves both correct functionality and enhanced performance. 
This motivates the necessity of systematically evaluating LLMs' NFR-aware code generation capabilities—such as by examining how prompt variations affect the resulting code.

In this paper, we introduce \tool, an automated framework that evaluates the robustness of LLMs in code generation while incorporating NFRs (i.e., NFR-aware code generation). \tool examines four commonly targeted dimensions of NFRs~\cite{rasheed2024aipoweredcodereviewllms} and considers three methodologies for evaluating LLMs in NFR-aware code generation.
Specifically, \tool examines NFR dimensions—including \textit{Code Design}, \textit{Reliability}, \textit{Readability}, and \textit{Performance}—using associated metrics to evaluate the generated code.
While considering these NFRs, \tool introduces the following methodology: 1) \textbf{\textit{Prompt Variations}}—to examine how generated code correctness and NFR quality change when users apply different prompts targeting specific NFR dimensions; 2) \textbf{\textit{Regression Testing}}—to investigate how an LLM’s NFR-aware code generation capabilities evolve after updates to the same model; and 3) \textbf{\textit{NFR-Aware Code Generation Workflows}}—to explore how different workflows for applying LLMs to address NFRs affect the resulting code. Through these NFR dimensions and methodologies, \tool is able to reveal potential robustness issues in LLMs when generating code with NFR awareness.

We apply \tool to three popular LLM families~\cite{minaee2024largelanguagemodelssurvey, zhao2025surveylargelanguagemodels} in our study: \textit{GPT-3.5-turbo} and \textit{GPT-4o} from OpenAI, and \textit{Claude-3.5} from Anthropic. Using four widely adopted coding benchmarks~\cite{huang2023agentcoder,lin2024soen101}—\textit{HumanEval}, \textit{HumanEval-ET}, \textit{MBPP}, and \textit{MBPP-ET}—we investigate the robustness of each model using \tool's evaluation methodology. \tool evaluates the robustness of LLMs' NFR-aware code generation capabilities by analyzing both functional correctness and NFR-related metrics (i.e., code smells, readability, exception handling, and execution time), along with their average values and STandard DEViations (STDEV). 
Our experimental results demonstrate that \tool effectively uncovers robustness issues in the tested LLMs through each methodology.
While code generation without considering NFRs (i.e., \raw) achieves stable Pass@1 scores (with an average STDEV of 0.48) across prompt variations, using different prompts to guide LLMs in addressing NFRs can reduce Pass@1 scores by up to 39\% compared to \raw. It also results in a significantly higher STDEV of 2.48, indicating greater variation and reduced robustness. Incorporating NFRs in the prompt generally helps improve NFR quality metrics. However, NFR metrics, especially related to readability and exception-handling, are more sensitive to prompt variation, where they have higher STDEV across prompts.

%While the models we studied were not robust to prompt variations, their 
We found that different versions of the LLM may introduce some trade-offs between code correctness and NFR metrics. 
For example, the newer version of \textit{GPT-4o} reduces code correctness in favor of improved performance (shorter execution time). However, not all LLMs exhibit clear trade-off patterns. For instance, \textit{GPT-3.5-turbo} shows completely opposite trends across the two benchmarks, \textit{HumanEval} and \textit{MBPP} compared to \textit{GPT-4o}. This inconsistency may make it difficult for users to understand the model's NFR-aware code generation capability or for developers to identify areas for improvement, highlighting potential robustness issues in LLMs. 
We also found robustness issues when changing the interaction workflows with the LLMs. Integrating NFRs and functional requirements into a single prompt (i.e., \first) generally produces code with higher correctness compared to asking the LLM to enhance \raw-generated code with the same NFRs (i.e., \second). Nevertheless, \second performs better in reducing code smells and improving readability, while \first is more effective at adding exception-handling statements and optimizing execution time. 
Overall, these findings suggest that robustness issues exist in the NFR-aware code generation capabilities of LLMs, highlighting the need for carefully selecting suitable prompts, model versions, and well-designed workflows during development.

We summarize our contributions as follows:

\begin{itemize}[noitemsep, topsep=3pt, leftmargin=*]
    \item We propose \tool, a novel framework for systematically evaluating the robustness of LLMs in addressing Non-Functional Requirements (NFRs) during code generation.
    \item We reveals potential robustness issues within popular LLM families (e.g., \textit{GPT} and \textit{Claude}). Their ability to generate NFR-aware code is significantly affected by prompt variations, model updates, and different workflows, impacting both code correctness and NFR metrics.
    \item Our comprehensive experiments emphasize the importance of establishing continuous quality assurance when integrating LLMs into real-world development. Frameworks like \tool can be used to monitor the robustness of deployed LLMs in real-world scenarios, preventing unexpected changes in LLM-based product behavior. 
    \item We provide the replication package and data to support reproduction and future studies~\cite{replication_package}.
\end{itemize}

\noindent{\textbf{Paper Organization}.} Section \ref{sec:related} reviews related work. Section \ref{sec:methodology} describes how \tool works. Section \ref{sec:evaluation} presents our experiment with \tool. Section \ref{sec:discussion} discusses the key findings and their implications. Section \ref{sec:threats} outlines potential threats to validity. Finally, Section \ref{sec:conclusion} concludes the paper.

\section{Related Work} \label{sec:related} 
\phead{Non-Functional Requirements (NFRs) in Coding.}
Prior studies proposed approaches for examining or refining existing source code to meet the NFRs.
% code smell / code design
\citet{pereira2022code} summarized a series of studies on detecting and visualizing code smells that negatively impact code design.
% readability
\citet{vitale2023using} trained models to improve the readability of given code snippets and \citet{li2023they} identified readability issues from logging code.
% reliability
\citet{zhang2020learning} proposed an automated approach to generate exception handling code based on existing source code to improve the overall software reliability.
% performance
\citet{biringa2023paceprogramanalysisframework} proposed a program analysis framework that provides continuous feedback on the performance impact of pending code updates, while \citet{zhao2023easyviewbringingperformanceprofiles} developed tools to help developers identify and resolve performance inefficiencies. All these studies highlight the importance of NFRs in real-world software development; however, they are limited to study only specific types of NFR tasks and a narrow set of metrics. In contrast, our work investigates how code quality changes across four NFR dimensions using multiple metrics, offering a more comprehensive aspects for studying code quality.

\phead{Exploring NFRs in Code Generation with LLMs.}
Several studies have explored the potential of LLMs in addressing NFRs to improve code quality across various dimensions. For instance, \citet{wu2024ismell} integrated LLMs with traditional code smell detection tools to automatically reduce code smells. \citet{xu2025mantraenhancingautomatedmethodlevel} leveraged LLMs to enhance code readability through automated refactoring. \citet{han2024archcodeincorporatingsoftwarerequirements} incorporated software requirements from textual descriptions to enable NFR-aware code generation, improving aspects such as reliability. \citet{gao2024search} utilized LLMs to optimize the execution efficiency of source code.
Unlike prior studies that focus on specific NFR-related tasks or refining existing code, our work emphasizes the inherent capabilities of LLMs in generating NFR-aware code. In addition, recent research has pointed out that addressing NFRs may negatively impact the functional correctness of generated code. For example, \citet{singhal2024nofunevalfunnycodelms} proposed a new benchmark to evaluate LLM in NFR-aware code generation and found that current models often struggle with such tasks. \citet{waghjale2024eccoimprovemodelgeneratedcode} studied how to improve code execution speed while preserving functional correctness. While these works focus on identifying or addressing NFR-related issues in LLM-generated code, our study examines how the NFR-aware code generation capabilities of LLMs vary across different usage scenarios and exploring the associated robustness challenges.

\phead{Studying the Robustness of LLMs in Code Generation.}
%Several studies have investigated the robustness of LLMs in code generation. 
\citet{wang2022recode}, \citet{chen2024nlperturbator}, and \citet{shirafuji2023exploring} explored robustness by perturbating different components in the prompts (e.g., problem descriptions, docstrings) with diverse patterns.
%Additionally, several studies have identified potential robustness issues that arise when LLMs are updated. 
\citet{chen2023chatgpt} and \citet{lin2024soen101} reported that ChatGPT's performance on code generation can change substantially between different versions of the same model. \citet{mishra2024granite} examined how robustness varies across various models and model sizes. 
These studies primarily focused on the functional correctness of the generated code. Given the critical role of NFRs in software development, our study addresses the importance of exploring the impact of incorporating NFR considerations into various coding workflows for LLM-based code generation. We also study the stability on the functional and NFR code quality across semantically equivalent prompts and model versions.  

\section{Methodology}\label{sec:methodology}

In this paper, we propose \tool, an automated framework that evaluates the robustness of LLMs when addressing Non-Functional Requirements (NFRs) in code generation. (i.e., \textit{NFR-aware code generation}). 
In this section, we provide details of \tool, including the four NFR dimensions along with their associated metrics, as well as the three complementary methodologies it incorporates. %Notably, \tool is extensible—it allows for the addition of new NFR dimensions or the use of alternative benchmarks beyond those used in this study.

\subsection{Overview of \tool}\label{sec:overview}

% https://docs.google.com/presentation/d/1s5D7iowsujq7Mr5XTLVv3j3WY8FbARtrGnIMFsudjQU/edit#slide=id.g3332ee6340c_0_0
\begin{figure*}
  \centering
  \includegraphics[width=\textwidth]{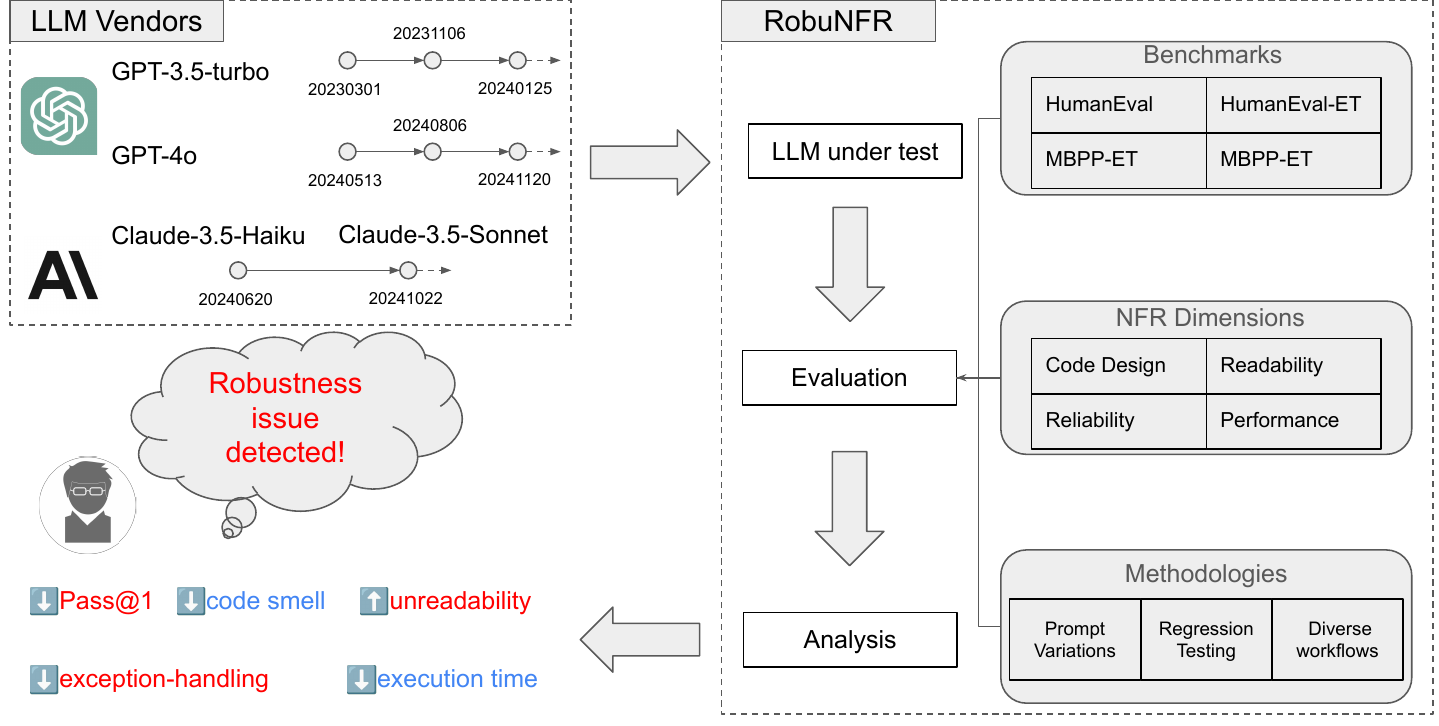}
  \caption{Overview of \tool. \tool leverages three methodologies to evaluate the code generation capabilities of LLM across NFR dimensions using various code benchmarks, aiming to reveal potential robustness issues in the LLM under test. } 
  ~\label{fig:overview}
  \vspace{-5mm}
\end{figure*}

NFRs, such as maintainability and readability, are critical aspects of code quality. However, existing studies on code generation often overlook NFRs and only focus on functional correctness metrics. 
For example, Pass@1~\cite{chen2021evaluating,chen2023chatgpt} is commonly used to assess whether the code generated by the LLM passes all test cases on its first attempt. 
However, without considering NFRs, the generated code might be only functionally correct but lack reliability, readability, or efficiency. Such neglect can lead to significant maintenance challenges and impact software quality~\cite{chung2009non}. 
When users leverage LLMs to address NFRs, they may do so in different ways—for example, by using varied prompts to express the same NFRs, by asking LLM to generate NFR-aware code, or by asking the LLM to improve existing code while considering NFRs. These different usage patterns can lead to different results~\cite{wang2022recode, chen2024nlperturbator,shirafuji2023exploring}. A robust LLM should produce consistent outputs when addressing the same NFRs, regardless of how the request is phrased or applied. However, measuring the robustness of LLMs in this context remains a challenging task. Hence, \tool is designed to study the capability and robustness of LLMs in addressing NFRs during the code generation process. 

As shown in Figure~\ref{fig:overview}, for the given LLM, \tool integrates existing coding benchmarks (e.g., HumanEval) with a set of specified NFR dimensions (e.g., Code Design) and employs three distinct evaluation approaches to generate comprehensive evaluation results. Specifically, our framework adopts three evaluation methodologies: 1) \textbf{\textit{Prompt Variations}}, assessing whether the functional and non-functional quality of LLM-generated code varies when using different prompts that convey the same NFRs; 2) \textbf{\textit{Regression Testing}}, evaluating differences in the NFR-aware code generation capabilities of an LLM after updates to the same model; and 3) \textbf{\textit{NFR-Aware Code Generation Workflows}}, analyzing the impact on how LLMs generate NFR-aware code when NFRs are addressed either through a one-shot prompt or sequentially. During the evaluation process, \tool provides a comprehensive set of metrics to assess both the functional correctness and code quality of the generated code that addresses NFRs. These metrics capture changes in LLM-generated code during evaluation, thereby highlighting potential robustness issues in NFR-aware code generation.

\subsection{Studied NFR Dimensions and Metrics}~\label{NFR:Metric}\label{metrics} 
In addition to functional correctness, \tool evaluates four NFR dimensions that contribute to a code's maintainability, reliability, and efficiency. Below, we describe functional correctness and each NFR dimension, along with the corresponding metric used for evaluation.

\phead{Functional Correctness} refers to whether the generated code satisfies the intended functionality. We use Pass@1 \cite{chen2021evaluating} to determine whether the code passes all test cases on its first attempt.

\noindent{\textbf{Code design}} refers to the structural and architectural quality of code, where bad designs can significantly hinder maintainability and scalability~\cite{walter2016relationship, fowler2018refactoring}. We use the \textit{Refactor} checker of Pylint~\cite{pylint_messages_refactor} to detect code smells. It includes predefined static code checkers to detect various code smells. We calculate and report the code smell density as the number of detected smells per 10 Lines Of Code (LOC) since the generated code may have different lengths.

\noindent{\textbf{Reliability}} is the code's ability to handle unexpected inputs and ensure stable execution under various scenarios (e.g., exception handling)~\cite{zhang2020learning, pham2000software}. In particular, we measure whether the generated code includes exception-handling mechanisms, such as try-catch blocks, to gain insight into how well the code anticipates and manages potential errors. We calculate exception density as the number of exception-handling statements per 10 lines of code, as this metric highlights the extent of error-handling logic~\cite{de2017revisiting}. 

\noindent{\textbf{Readability}} is how easily code can be understood and modified. Readable code should follow coding style guidelines and conventions to ease understanding and collaboration~\cite{piantadosi2020does}. Similar to code design, we use use the \textit{Convention} checker of Pylint~\cite{pylint_messages_convention} to detect issues like inconsistent naming, incorrect indentation, and missing comments. We also report the density of readability issues per 10 lines of code.

\noindent{\textbf{Performance}} assesses the efficiency of code, where performance issues (e.g., slower execution) can cause higher operational costs and reduce user satisfaction~\cite{malik2013automatic}. We measure the execution time in milliseconds for all tests associated with each coding problem. To minimize measurement fluctuations, we run each test case five times and calculate the mean.

\subsection{Evaluation Methodology 1: Prompt Variations}\label{section:prompts}

\begin{table*}[t]
    \centering
        \caption{LLM generated prompt templates to consider non-functional requirements in code generation.}

    \scalebox{0.47}{
    \begin{tabular}{l|l|l|l}
    \toprule
    \cellcolor{red!15}Error-handle & \cellcolor{green!15}Code Smell & \cellcolor{blue!15}Readability & \cellcolor{yellow!15}Performance \\
    \midrule
       Incorporate various \textit{\underline{error handling}} techniques & Investigate various strategies to handle \textit{\underline{code smell}} & Evaluate different coding practices for \textit{\underline{readability}} & Optimize for \textit{\underline{performance}} \\
       Implement multiple \textit{\underline{exception handling}} strategies  & Minimize \textit{\underline{code smell}} & Investigate various techniques to enhance \textit{\underline{readability}} & Focus on enhancing \textit{\underline{performance}} \\
       Apply different \textit{\underline{error handling}} mechanisms & Eliminate \textit{\underline{code smell}} & Improve the code \textit{\underline{readability}} & Ensure the code \textit{\underline{runs efficiently}} \\
       Investigate different methods of \textit{\underline{managing exceptions}} & Identify and address different \textit{\underline{code smells}} & Ensure the code is \textit{\underline{readable}} & Prioritize \textit{\underline{runtime optimization}} \\
       Integrate diverse \textit{\underline{error handling}} approaches & Apply best practices to reduce \textit{\underline{code smell}} & Apply coding practices that enhance \textit{\underline{readability}} & Keep \textit{\underline{performance}} in mind while solving \\
       Utilize multiple \textit{\underline{error management}} techniques & Mitigate \textit{\underline{code smell}} & Focus on \textit{\underline{readability}} & Aim for \textit{\underline{high-performance}} execution \\
       Experiment with various ways to \textit{\underline{handle exceptions}}  & Tackle different \textit{\underline{code smell}} issues & Enhance the \textit{\underline{readability}} of the code & Reduce \textit{\underline{computational overhead}}\\
       Combine different \textit{\underline{error handling}} practices  & Implement techniques to prevent \textit{\underline{code smell}} & Implement strategies to make the code more \textit{\underline{readable}} & Emphasize \textit{\underline{speed and efficiency}} \\
       Evaluate multiple \textit{\underline{exception management}} strategies & Resolve \textit{\underline{code smell}} problems & Optimize the code for better \textit{\underline{readability}} & Ensure minimal \textit{\underline{resource consumption}} \\
       Develop a range of \textit{\underline{error handling}} solutions & Optimize code to avoid \textit{\underline{code smell}} & Adopt coding practices for improved \textit{\underline{readability}} & Maximize \textit{\underline{performance}} in your solution \\
    \bottomrule
    \end{tabular}
    } 
    %\vspace{-3mm}
    \label{tab:generated_prompt}
\end{table*}

%\phead{NFR-Aware Coding Prompt Templates.}
Prior research~\cite{chen2024nlperturbator, wang2022recode, shirafuji2023exploring} suggests that variations in prompt templates, even when preserving semantic context, can generate significantly different code. Hence, we repeat the code generation process using different but semantically equivalent prompts. To mitigate potential biases introduced by manually altering the prompts, we leverage GPT-4o-mini to generate various prompts for each dimension of NFRs while preserving the same semantics. This approach allows us to evaluate the stability of the results by measuring variations across different prompt templates, thereby clarifying how changes in prompts influence the code generated by LLMs. In other words, if the results remain stable, it indicates that the LLM is robust to variations in prompts.

\phead{Constructing Diverse NFR-Aware Prompts.}
Table \ref{tab:generated_prompt} shows the semantically equivalent prompt templates generated for each dimension of NFRs. Initially, we manually crafted a seed prompt with the structure: ``\textit{Consider [NFR] and complete the following code}'', where ``\textit{[NFR]}'' corresponds to specific non-functional requirements, such as code design or readability. We then provided the seed prompt to ChatGPT to generate 10 semantically equivalent prompts for the experiment.  
We incorporate all 10 prompt variants to assess the robustness of the LLM against semantic preserving changes in the prompt. In total, we execute NFR-aware code generation 40 times (10 variations per NFR) for each workflow and each LLM version. Although our experiments are conducted on existing code generation benchmarks (e.g., HumanEval and MBPP), \tool is highly adaptable. Future studies can easily tailor its process to accommodate new NFRs and additional benchmarks.

\subsection{Evaluation Methodology 2: Regression Testing}\label{sec:regression}
Previous research has shown that robustness issues exist in LLMs during version updates~\cite{chen2023chatgpt,lin2024soen101}. Although developers may claim that an LLM's code generation ability remains consistent after an update, the actual results often vary depending on the coding benchmark used. This discrepancy becomes even more pronounced in NFR-aware code generation tasks, where robustness issues may be more deeply concealed.

To address this challenge, \tool leverages the concept of regression testing to monitor robustness issues, specifically in NFR-aware code generation tasks. In particular, \tool incorporates several key components of regression testing:

\noindent{\textbf{Fixed Test Suite and Metrics:}} \tool evaluates both the older and newer versions using the same test suites (e.g., code generation benchmark) and metrics, guaranteeing a fair comparison.

\noindent{\textbf{Baseline Establishment:}} We define both the \raw (i.e., requirements contain only functional features without any additional NFRs) and the older model version as baselines, enabling a direct comparison of their behaviors.

\noindent{\textbf{Impact Analysis:}} By maintaining consistent evaluation criteria, \tool tracks trends in NFR-aware code generation and quantifies the extent of change.

\tool employs regression testing to reliably identify robustness issues during model updates. In other words, if a newer version of an LLM demonstrates consistent NFR-aware code generation capabilities across all regression test cases, it can be considered robust with respect to version changes.

\subsection{Evaluation Methodology 3: NFR-Aware Code Generation Workflows}\label{sec:approach}

\begin{figure*}
  \centering
  \includegraphics[width=1\textwidth]{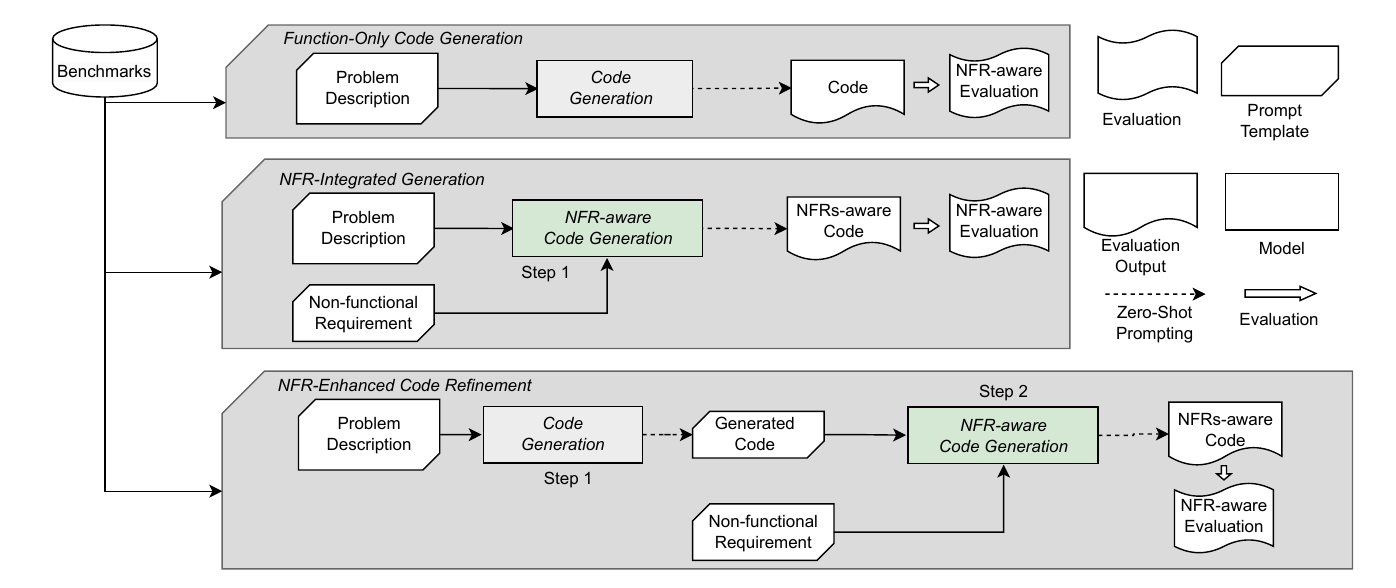}
  \caption{\tool defines three workflows as part of its NFR-aware code generation evaluation methodology. These workflows include Function-Only code generation, NFR-Integrated code generation, and NFR-Enhanced code refinement. We compare the functional and non-functional quality of the generated code across these workflows.}
  ~\label{fig:workflow}
  % \vspace{-5mm}
\end{figure*}

In addition to generating functional code, modern code generation tools, such as Cursor~\cite{cursor} and GitHub Copilot~\cite{copilotwebsite}, provide two typical workflows for NFR-aware code generation~\cite{copilotwebsite2}. 1) \textit{\textbf{NFR-integrated code generation}} involves developers providing both the functional and non-functional requirements in one prompt to generate the complete code in one shot. 
2) \textit{\textbf{NFR-enhanced code refinement}} refers to the process by which developers use an LLM to refine existing code, thereby enhancing code quality and better aligning it with specific requirements~\cite{white2024chatgpt}.
Figure \ref{fig:workflow} provides an overview of these workflows. The baseline workflow considers only the functional requirement (i.e., \textit{Functional-Only Code Generation}, denoted as \functional, and two NFR-aware code generation workflows (i.e., \first and \second).

While both workflows provide the instruction to generate code that satisfies specific requirements, the final output may be different, as the ways of interacting with LLMs may significantly affect the generated results~\cite{lee2024exploring}. Therefore, in \tool, we consider both of these two workflows to incorporate the four NFRs into the code. In the remainder of the paper, we denote NFR-integrated code generation as \first and NFR-aware code enhancement as \second for conciseness. 
To analyze the results, we compare the functional and non-functional code quality metrics across the code generated by three distinct workflows, examining the impact of NFR-aware code generation on overall code quality.

Figure~\ref{fig:promt_template} shows the prompt templates for each workflow. 
\functional only contains the functional requirement in the prompt. 
\first incorporates NFRs directly into the prompt template. For example, when considering reliability, the prompt asks the LLM to generate code that meets the functional requirements and optimize reliability in a single prompt. \second adopts a two-step process. It leverages the code generated by \functional, and it sends a separate prompt asking the LLM to enhance the code by addressing a specific NFR. 
Both \first and \second use the same NFR-aware prompt templates outlined in Table~\ref{tab:generated_prompt}.

\begin{figure}
    %\scriptsize
    \greybox{  
    \textbf{\underline{\ding{226}Function-Only Code Generation}}
    
\textit{Complete the following code.
        \\
        \#\# Input:
        \textcolor{darkgray}{`\{Problem Description\}'}
        \\
        \#\# Response:
        \textcolor{blue}{`\{Code\}'} \\
}

    \textbf{\underline{\ding{226}NFR-Integrated Code Generation}}

\textit{Given the problem description, generate code by considering \textcolor{red}{\{NFR\}}.''
        \\
        \#\# Input:
        \textcolor{darkgray}{`\{Problem Description\}'}
        \\
        \#\# Response:
        \textcolor{purple}{`\{NFR-aware Code\}'} \\
   }      
        \\
     \textbf{\underline{\ding{226}NFR-Enhanced Code Refinement}}\\
     \textbf{Step 1 - Existing code to be refined}  -> \textcolor{blue}{`\{Code\}'} \\
     \textbf{Step 2 - Refine the code with NFR }

\textit{Given the following code, your goal is to improve its \textcolor{red}{\{NFR\}}.''
        \\
        \#\# Input:
        \textcolor{blue}{`\{Code\}'}
        \\
        \#\# Response: 
        \textcolor{purple}{`\{NFR-aware Code\}'} 
} 
        }
    \caption{A simplified example of a prompt template for NFR-Aware code generation workflows.}
    \label{fig:promt_template}
\end{figure}

\section{Evaluation}\label{sec:evaluation}\label{setup}
Although \tool is not limited to specific benchmarks or LLMs under evaluation, in our study, we selected several popular models and coding benchmarks to demonstrate how \tool can reveal robustness issues in the NFR-aware code generation capabilities of LLMs.

\noindent{\textbf{Studied LLMs.}} We selected two widely recognized LLM families mentioned in recent LLM-focused surveys~\cite{minaee2024largelanguagemodelssurvey, zhao2025surveylargelanguagemodels}. Specifically, we evaluated \textit{GPT-3.5-turbo} and \textit{GPT-4o} from OpenAI, as well as \textit{Claude-3.5-Sonnet} and \textit{Claude-3.5-Haiku} from Anthropic, as these models allow for a comparison of different LLMs within the same family, sharing similar architectures and development processes. To support regression testing, we included two released versions for each LLM (e.g., \textit{gpt-3.5-turbo-1106} and \textit{gpt-3.5-turbo-0125} for \textit{GPT-3.5-turbo}). All interactions with the models were conducted via vendor-provided APIs. To minimize output variance and ensure deterministic responses, we set the temperature parameter to 0.

\noindent{\textbf{Benchmark Datasets.}} \tool can be applied to any code generation benchmark across different programming languages. In our study, we selected four commonly used datasets: HumanEval, HumanEval-ET, MBPP (Mostly Basic Python Programming), and MBPP-ET. These benchmarks are widely adopted in code generation research~\cite{huang2023agentcoder,lin2024soen101} and include test cases for evaluating the functional correctness of generated code. HumanEval~\citep{chen2021evaluating} comprises 164 programming problems, while MBPP~\citep{austin2021program} includes 427 programming problems (we used the sanitized version provided by the original authors). Furthermore, HumanEval-ET and MBPP-ET, published by \citet{dong2023codescore}, use the same problems as HumanEval and MBPP but offer more test cases with approximately 100 test cases for each problem. 

\noindent{\textbf{Environment.}} Our experiments were conducted on a Mac Mini (Apple M4, 10 cores, 16GB RAM), using Python 3.9.19 to implement \tool and the evaluation scripts. The OpenAI API library used was version 1.14.3, and the Claude API library used was version 0.39.0. For detecting code smells and readability issues, we used Pylint version 3.2.5. We have made our framework code and evaluation data publicly available to support future research~\cite{replication_package}. 

\subsection*{RQ1: How do variations in prompts affect the robustness of LLMs in NFR-aware code generation?}

\label{sec:rq1}

\noindent{\underline{\textit{Motivation.}}} 
Different users may express the same NFRs using diverse prompts when instructing an LLM to generate code. This research question explores how variations in prompt wording influence the LLMs in producing code that adheres to NFRs and functional correctness.

\noindent{\underline{\textit{Approach.}}} 
We adopt Evaluation Methodology 1 (i.e., Prompt Variations) as discussed in Section~\ref{section:prompts}. We compute the Pass@1 and NFR metrics for generating four types of NFR-aware code—design, reliability, readability, and performance—using different prompts. Pass@1 measures the functional correctness of the code by evaluating whether the first generated solution successfully passes all the provided tests~\cite{chen2021evaluating}. A Pass@1 of 100 means the first generated code can pass all the tests in 100\% of the coding problems in a benchmark. 
Additionally, we compare the changes in average (AVG) and standard deviation (STDEV) relative to our baseline (\raw), where code is generated considering only functional requirements. We then extend our study across various models and four benchmarks, as discussed in Section~\ref{setup}.

\noindent{\underline{\textit{Results.}}}
\textit{\textbf{Integrating NFRs almost always reduces Pass@1 scores across all models and benchmarks by up to 39\%. Notably, some models appear more susceptible to the challenges posed by NFR dimensions.}} Table~\ref{tab:rq1_pass1} presents the results for the baseline (\raw) and all four NFR dimensions using different prompts that convey the same meaning. The Pass@1 scores for the NFR dimensions are mostly lower than that of the baseline, particularly in the Reliability dimension, where the Claude-3.5-Haiku model shows a 39\% reduction in Pass@1. Additionally, we observed that the average Pass@1 score dropped by 9.76\% between the Claude-3.5-Sonnet and Claude-3.5-Haiku models, compared to a 3.45\% drop between GPT-3.5-turbo and GPT-4o. This type of comparison—between two generations of the Claude family and two generations of the GPT family—highlights potential directions for improving model architectures and training processes. Our results show that models from the Claude family tend to be more adversely affected by the additional challenges introduced by the NFRs.

% https://docs.google.com/spreadsheets/d/1KOLnYyaQ5jpL9ybrgJW4UQZmbXFFmCrkYBIsRnVi88c/edit?gid=1068127053#gid=1068127053
\begin{table}[t]
\setlength{\tabcolsep}{2pt}
\centering

\caption{The Pass@1 column represents the Pass@1 scores, along with their STDEV across 10 semantically equivalent prompts. $\Delta$ indicates the percentage difference in Pass@1 of the same model version between the NFR-aware results and the Function-Only result.}

\scalebox{0.7}{
\begin{tabular}{ll|rrr|rrr|rrr|rrrr}\toprule
\multirow{2}{*}{NFR Dimension} &\multirow{2}{*}{Model} &\multicolumn{3}{c}{HumanEval} &\multicolumn{3}{c}{HumanEval-ET} &\multicolumn{3}{c}{MBPP} &\multicolumn{3}{c}{MBPP-ET} \\\cmidrule{3-14}
& &Pass@1 &STDEV &$\Delta$(\%) &Pass@1 &STDEV &$\Delta$(\%) &Pass@1 &STDEV &$\Delta$(\%) &Pass@1 &STDEV &$\Delta$(\%) \\\midrule

\multirow{4}{*}{Function-Only} 
  & GPT-3.5-turbo         & 72.50  & 0.73 & -                              & 64.33 & 1.05 & -                              & 67.82 & 0.48 & -                              & 47.21 & 0.39 & - \\
  & GPT-4o                & 90.55  & 1.16 & -                              & 80.18 & 0.96 & -                              & 74.43 & 0.55 & -                              & 53.37 & 0.34 & - \\
  & Claude-3.5-Sonnet     & 89.39  & 0.33 & -                              & 78.54 & 0.51 & -                              & 75.97 & 0.27 & -                              & 54.94 & 0.13 & - \\
  & Claude-3.5-Haiku      & 86.22  & 0.33 & -                              & 75.61 & 0.43 & -                              & 72.37 & 0.00 & -                              & 52.93 & 0.00 & - \\
\midrule
\multirow{4}{*}{Code Design} 
  & GPT-3.5-turbo         & 72.87  & 1.82 & 0.51                           & 64.51 & 2.15 & 0.28                          & 67.68 & 1.40 & \textcolor{red}{$\downarrow$ 0.21} & 47.61 & 1.08 & 0.85 \\
  & GPT-4o                & 89.33  & 0.77 & \textcolor{red}{$\downarrow$ 1.35} & 79.63 & 1.08 & \textcolor{red}{$\downarrow$ 0.69} & 73.37 & 1.12 & \textcolor{red}{$\downarrow$ 1.42} & 52.58 & 1.10  & \textcolor{red}{$\downarrow$ 1.48} \\
  & Claude-3.5-Sonnet     & 84.02  & 1.85 & \textcolor{red}{$\downarrow$ 6.01} & 72.56 & 1.93 & \textcolor{red}{$\downarrow$ 7.61} & 70.87 & 2.58 & \textcolor{red}{$\downarrow$ 6.71} & 49.79 & 2.23 & \textcolor{red}{$\downarrow$ 9.37} \\
  & Claude-3.5-Haiku      & 80.73  & 2.42 & \textcolor{red}{$\downarrow$ 6.37} & 70.12 & 2.62 & \textcolor{red}{$\downarrow$ 7.26} & 64.73 & 5.47 & \textcolor{red}{$\downarrow$ 10.56} & 45.67 & 3.93 & \textcolor{red}{$\downarrow$ 13.72} \\
\midrule

\multirow{4}{*}{Readability} 
  & GPT-3.5-turbo         & 73.17  & 2.80 & 0.92                           & 64.33 & 1.91 & 0.00                          & 68.76 & 1.51 & 1.39                           & 48.41 & 1.19 & 2.54 \\
  & GPT-4o                & 91.40  & 1.64 & 0.94                           & 80.98 & 1.60 & 1.00                          & 75.04 & 0.87 & 0.82                           & 53.63 & 0.89 & 0.49 \\
  & Claude-3.5-Sonnet     & 86.46  & 1.80 & \textcolor{red}{$\downarrow$ 3.28} & 75.85 & 1.81 & \textcolor{red}{$\downarrow$ 3.43} & 73.35 & 2.64 & \textcolor{red}{$\downarrow$ 3.45} & 51.66 & 1.52 & \textcolor{red}{$\downarrow$ 5.97} \\
  & Claude-3.5-Haiku      & 84.02  & 3.41 & \textcolor{red}{$\downarrow$ 2.55} & 73.78 & 3.26 & \textcolor{red}{$\downarrow$ 2.42} & 61.55 & 7.79 & \textcolor{red}{$\downarrow$ 14.95} & 44.31 & 5.41 & \textcolor{red}{$\downarrow$ 16.29} \\
\midrule

\multirow{4}{*}{Reliability} 
  & GPT-3.5-turbo         & 68.29  & 3.50 & \textcolor{red}{$\downarrow$ 5.81}  & 59.09 & 3.62 & \textcolor{red}{$\downarrow$ 8.15}  & 42.93 & 13.93& \textcolor{red}{$\downarrow$ 36.70}  & 29.46 & 9.60 & \textcolor{red}{$\downarrow$ 37.6} \\
  & GPT-4o                & 88.29  & 1.25 & \textcolor{red}{$\downarrow$ 2.50}  & 76.22 & 1.52 & \textcolor{red}{$\downarrow$ 4.94}  & 71.59 & 0.88 & \textcolor{red}{$\downarrow$ 3.82}  & 50.02 & 0.70 & \textcolor{red}{$\downarrow$ 6.28} \\
  & Claude-3.5-Sonnet     & 81.83  & 2.64 & \textcolor{red}{$\downarrow$ 8.46}  & 70.12 & 2.96 & \textcolor{red}{$\downarrow$ 10.72} & 69.32 & 1.81 & \textcolor{red}{$\downarrow$ 8.75}  & 46.79 & 1.96 & \textcolor{red}{$\downarrow$ 14.83} \\
  & Claude-3.5-Haiku      & 73.05  & 2.26 & \textcolor{red}{$\downarrow$ 15.27} & 62.20 & 2.02 & \textcolor{red}{$\downarrow$ 17.74} & 47.45 & 4.21 & \textcolor{red}{$\downarrow$ 34.43} & 32.04 & 2.68 & \textcolor{red}{$\downarrow$ 39.47} \\
\midrule

\multirow{4}{*}{Performance} 
  & GPT-3.5-turbo         & 70.79  & 3.45 & \textcolor{red}{$\downarrow$ 2.36}  & 61.83 & 2.93 & \textcolor{red}{$\downarrow$ 3.89}  & 66.63 & 1.92 & \textcolor{red}{$\downarrow$ 1.75}  & 47.82 & 1.47 & 1.29 \\
  & GPT-4o                & 89.33  & 2.12 & \textcolor{red}{$\downarrow$ 1.35}  & 80.18 & 1.61 & 0.00                              & 74.07 & 0.72 & \textcolor{red}{$\downarrow$ 0.48}  & 53.56 & 0.80 & 0.36 \\
  & Claude-3.5-Sonnet     & 83.29  & 1.53 & \textcolor{red}{$\downarrow$ 6.82}  & 74.02 & 1.40 & \textcolor{red}{$\downarrow$ 5.76}  & 72.04 & 1.76 & \textcolor{red}{$\downarrow$ 5.17}  & 51.43 & 2.08 & \textcolor{red}{$\downarrow$ 6.39} \\
  & Claude-3.5-Haiku      & 81.32  & 1.98 & \textcolor{red}{$\downarrow$ 4.81}  & 71.95 & 0.96 & \textcolor{red}{$\downarrow$ 4.84}  & 70.73 & 2.36 & \textcolor{red}{$\downarrow$ 2.27}  & 49.41 & 2.39 & \textcolor{red}{$\downarrow$ 6.65} \\

\bottomrule
\end{tabular}
}
\label{tab:rq1_pass1}
\vspace{-3mm}
\end{table}

{\textit{\textbf{Compared to the baseline (\raw), all models exhibit consistently higher STDEV values in Pass@1 when handling NFR dimensions, indicating increased robustness issues.}}} As shown in Table~\ref{tab:rq1_pass1}, the STDEV for \raw's Pass@1 is generally lower than that for NFR dimensions across all models and benchmarks. For example, the STDEV for \raw's Pass@1 is between 0.00 to 1.16 across all the models and benchmarks, while the STDEV, when considering Code Design, is between 0.77 to 5.47. This suggests that when users employ different wordings in prompts to convey the same meaning in NFR dimensions, the Pass@1 of the LLMs varies more than in \raw dimensions. {\textit{\textbf{Additionally, our analysis indicates that different NFR dimensions exhibit varying levels of robustness issues.}} In particular, the Reliability dimension presents the largest variance in certain models. For instance, as shown in Table~\ref{tab:rq1_pass1}, GPT-3.5-turbo reaches a peak STDEV of 13.93 across all four NFR dimensions, while Claude-3.5-Haiku achieves a peak STDEV of 4.21. These results suggest that certain prompts can significantly impact code correctness for these models, indicating that less robust models may be more sensitive to variations in Pass@1 scores under the reliability dimension.

% https://docs.google.com/spreadsheets/d/1KOLnYyaQ5jpL9ybrgJW4UQZmbXFFmCrkYBIsRnVi88c/edit?gid=2014524517#gid=2014524517
\begin{table}[t]
\centering

\caption{Columns \textit{code smell density}, \textit{unreadability density}, \textit{exception-handling density}, and \textit{execution time (millisecond)} represent the NFR metrics (Section~\ref{metrics}). Each metric includes standard deviations and $\Delta$\%, which indicates the percentage difference between NFR-aware results and \raw results. }

\setlength{\tabcolsep}{2pt}
\scalebox{0.49}{

\begin{tabular}{ll|rrrr|rrrr}\toprule
\multirow{2}{*}{NFR Dimension} & \multirow{2}{*}{Model} & \multicolumn{4}{c}{HumanEval} & \multicolumn{4}{c}{MBPP} \\\cmidrule{3-10}
 &  & code smell($\Delta$\%) & unreadability($\Delta$\%) & exception-handling($\Delta$\%) & execution time($\Delta$\%) & code smell($\Delta$\%) & unreadability($\Delta$\%) & exception-handling($\Delta$\%) & execution time($\Delta$\%) \\\midrule
\multirow{4}{*}{Function-Only} 
    & GPT-3.5-turbo      & 0.31±0.01        & 3.42±0.04        & 0.036±0.003      & 112.63±48.09     & 0.27±0.01        & 3.44±0.03        & 0.011±0.000      & 43.18±7.96 \\
    & GPT-4o             & 0.12±0.01        & 2.62±0.03        & 0.026±0.005      & 75.92±4.23       & 0.12±0.00        & 3.18±0.02        & 0.130±0.005      & 37.23±1.37 \\
    & Claude-3.5-Sonnet  & 0.10±0.01        & 2.03±0.01        & 0.037±0.003      & 57.06±2.93       & 0.08±0.00        & 2.67±0.01        & 0.069±0.002      & 40.92±3.68 \\
    & Claude-3.5-Haiku   & 0.06±0.00        & 2.60±0.02        & 0.022±0.000      & 70.31±13.75      & 0.03±0.00        & 2.69±0.00        & 0.041±0.000      & 34.40±0.30 \\\midrule
\multirow{4}{*}{Code Design} 
    & GPT-3.5-turbo      & 0.22±0.03 (\textcolor{red}{$\downarrow$29.0})  & 2.68±0.20 (\textcolor{red}{$\downarrow$21.6})  & 0.051±0.038 (41.7)  & 92.93±32.03 (\textcolor{red}{$\downarrow$17.49})  & 0.28±0.03 (3.7)   & 4.68±0.48 (36.0)  & 0.117±0.106 (963.6)  & 58.78±11.12 (36.13) \\
    & GPT-4o             & 0.06±0.00 (\textcolor{red}{$\downarrow$50.0})  & 1.27±0.15 (\textcolor{red}{$\downarrow$51.5})  & 0.091±0.018 (250.0) & 81.22±6.32 (6.98)  & 0.05±0.00 (\textcolor{red}{$\downarrow$58.3})  & 1.79±0.08 (\textcolor{red}{$\downarrow$43.7})  & 0.363±0.029 (179.2) & 38.30±3.03 (2.87) \\
    & Claude-3.5-Sonnet  & 0.02±0.01 (\textcolor{red}{$\downarrow$80.0})  & 0.79±0.10 (\textcolor{red}{$\downarrow$61.1})  & 0.148±0.064 (300.0) & 47.27±15.87 (\textcolor{red}{$\downarrow$17.16})  & 0.03±0.01 (\textcolor{red}{$\downarrow$62.5})  & 1.66±0.18 (\textcolor{red}{$\downarrow$37.8})  & 0.454±0.142 (558.0) & 36.42±0.18 (\textcolor{red}{$\downarrow$11.00}) \\
    & Claude-3.5-Haiku   & 0.02±0.01 (\textcolor{red}{$\downarrow$66.7})  & 1.54±0.50 (\textcolor{red}{$\downarrow$40.8})  & 0.176±0.118 (700.0) & 55.95±1.23 (\textcolor{red}{$\downarrow$20.42})  & 0.01±0.01 (\textcolor{red}{$\downarrow$66.7})  & 1.95±0.64 (\textcolor{red}{$\downarrow$27.5})  & 0.445±0.171 (985.4) & 35.18±1.83 (2.27) \\\midrule
\multirow{4}{*}{Readability} 
    & GPT-3.5-turbo      & 0.18±0.03 (\textcolor{red}{$\downarrow$41.9})  & 2.47±0.24 (\textcolor{red}{$\downarrow$27.8})  & 0.016±0.005 (\textcolor{red}{$\downarrow$55.6})  & 120.41±49.29 (6.91)  & 0.24±0.02 (\textcolor{red}{$\downarrow$11.1})  & 4.14±0.49 (20.3)  & 0.013±0.006 (18.2)  & 49.45±10.22 (14.52) \\
    & GPT-4o             & 0.06±0.01 (\textcolor{red}{$\downarrow$50.0})  & 1.25±0.07 (\textcolor{red}{$\downarrow$52.3})  & 0.029±0.007 (11.5)   & 84.49±5.31 (11.29)   & 0.07±0.01 (\textcolor{red}{$\downarrow$41.7})  & 1.86±0.19 (\textcolor{red}{$\downarrow$41.5})  & 0.107±0.033 (\textcolor{red}{$\downarrow$17.7})  & 36.39±2.33 (\textcolor{red}{$\downarrow$2.26}) \\
    & Claude-3.5-Sonnet  & 0.04±0.02 (\textcolor{red}{$\downarrow$60.0})  & 0.97±0.20 (\textcolor{red}{$\downarrow$52.2})  & 0.084±0.042 (127.0)  & 44.04±13.54 (\textcolor{red}{$\downarrow$22.82})  & 0.05±0.02 (\textcolor{red}{$\downarrow$37.5})  & 1.54±0.24 (\textcolor{red}{$\downarrow$42.3})  & 0.261±0.097 (278.3)  & 39.04±8.94 (\textcolor{red}{$\downarrow$4.59}) \\
    & Claude-3.5-Haiku   & 0.02±0.01 (\textcolor{red}{$\downarrow$66.7})  & 1.38±0.20 (\textcolor{red}{$\downarrow$46.9})  & 0.088±0.040 (300.0)  & 61.56±10.61 (\textcolor{red}{$\downarrow$12.44})  & 0.02±0.01 (\textcolor{red}{$\downarrow$33.3})  & 1.61±0.10 (\textcolor{red}{$\downarrow$40.1})  & 0.226±0.086 (451.2)  & 35.33±1.56 (2.70) \\\midrule
\multirow{4}{*}{Reliability} 
    & GPT-3.5-turbo      & 0.34±0.10 (9.7)   & 2.81±0.40 (\textcolor{red}{$\downarrow$17.8})  & 1.342±0.247 (3627.8) & 117.86±41.17 (4.64)  & 0.36±0.07 (33.3)  & 3.25±0.62 (\textcolor{red}{$\downarrow$5.5})   & 1.601±0.222 (14454.5) & 40.96±0.67 (\textcolor{red}{$\downarrow$5.14}) \\
    & GPT-4o             & 0.10±0.04 (\textcolor{red}{$\downarrow$16.7})  & 1.45±0.16 (\textcolor{red}{$\downarrow$44.7})  & 0.942±0.157 (3523.1) & 90.48±4.98 (19.18)   & 0.17±0.08 (41.7)  & 2.61±0.19 (\textcolor{red}{$\downarrow$17.9})  & 1.584±0.204 (1118.5)  & 35.09±0.27 (\textcolor{red}{$\downarrow$5.75}) \\
    & Claude-3.5-Sonnet  & 0.05±0.01 (\textcolor{red}{$\downarrow$50.0})  & 1.07±0.05 (\textcolor{red}{$\downarrow$47.3})  & 1.177±0.152 (3081.1) & 53.22±7.62 (\textcolor{red}{$\downarrow$6.73})  & 0.05±0.01 (\textcolor{red}{$\downarrow$37.5})  & 1.98±0.45 (\textcolor{red}{$\downarrow$25.8})  & 1.354±0.107 (1862.3)  & 43.66±15.62 (6.70) \\
    & Claude-3.5-Haiku   & 0.03±0.01 (\textcolor{red}{$\downarrow$50.0})  & 1.76±0.16 (\textcolor{red}{$\downarrow$32.3})  & 1.006±0.079 (4472.7)  & 81.29±41.04 (15.62)  & 0.01±0.00 (\textcolor{red}{$\downarrow$66.7})  & 1.48±0.20 (\textcolor{red}{$\downarrow$45.0})  & 1.115±0.075 (2619.5)  & 34.69±0.43 (\textcolor{red}{$\downarrow$0.84}) \\\midrule
\multirow{4}{*}{Performance} 
    & GPT-3.5-turbo      & 0.27±0.04 (\textcolor{red}{$\downarrow$12.9})  & 3.21±0.22 (\textcolor{red}{$\downarrow$6.1})   & 0.016±0.005 (\textcolor{red}{$\downarrow$55.6})  & 63.48±33.85 (\textcolor{red}{$\downarrow$43.64})  & 0.28±0.06 (3.7)   & 5.99±0.30 (74.1)  & 0.011±0.002 (0.0)   & 51.61±10.93 (19.52) \\
    & GPT-4o             & 0.08±0.00 (\textcolor{red}{$\downarrow$33.3})  & 1.64±0.13 (\textcolor{red}{$\downarrow$37.4})  & 0.027±0.010 (3.8)   & 74.34±1.00 (\textcolor{red}{$\downarrow$2.08})   & 0.12±0.02 (0.0)   & 3.26±0.18 (2.5)   & 0.103±0.026 (\textcolor{red}{$\downarrow$20.8})  & 34.50±0.43 (\textcolor{red}{$\downarrow$7.33}) \\
    & Claude-3.5-Sonnet  & 0.05±0.02 (\textcolor{red}{$\downarrow$50.0})  & 1.67±0.11 (\textcolor{red}{$\downarrow$17.7})  & 0.028±0.005 (\textcolor{red}{$\downarrow$24.3})  & 35.20±1.72 (\textcolor{red}{$\downarrow$38.31})  & 0.05±0.00 (\textcolor{red}{$\downarrow$37.5})  & 2.51±0.12 (\textcolor{red}{$\downarrow$6.0})   & 0.096±0.047 (39.1)  & 34.62±0.64 (\textcolor{red}{$\downarrow$15.40}) \\
    & Claude-3.5-Haiku   & 0.02±0.01 (\textcolor{red}{$\downarrow$66.7})  & 2.33±0.08 (\textcolor{red}{$\downarrow$10.4})  & 0.033±0.008 (50.0)   & 98.39±24.43 (39.94)  & 0.02±0.00 (\textcolor{red}{$\downarrow$33.3})  & 2.42±0.11 (\textcolor{red}{$\downarrow$10.0})  & 0.070±0.027 (70.7)  & 37.19±2.53 (8.11) \\
\bottomrule
\end{tabular}
}
\label{tab:rq1_nfr}
\vspace{-3mm}
\end{table}

{\textit{\textbf{Incorporating NFRs into prompts generally improves code quality by reducing code smells and enhancing readability. }}} As shown in Table~\ref{tab:rq1_nfr}, incorporating NFRs generally reduces code smells—achieving an average reduction of 34.93\% compared to \raw—and enhance readability, with an average improvement of 24.32\% across all models and benchmarks. We also observed that the extent of these improvements varies across models. \textbf{\textit{Claude models consistently generating code with fewer smells and better readability compared to GPT models.}} For instance, in the Code Design dimension, Claude-3.5-Sonnet and Claude-3.5-Haiku achieved code smell density values of 0.02 and 0.01, compared to GPT-3.5-turbo and GPT-4o at 0.22 and 0.06. Although all models improve over the baseline, Claude models show a more pronounced reduction in code smell—a trend consistent across all NFR dimensions and benchmarks, especially in Readability dimension. However, differences in the Reliability and Performance dimension are less distinct.

{\textit{\textbf{Incorporating NFRs into prompts results in higher STDEV values for the code smell, unreadability, and exception-handling metrics, suggesting that different prompts introduce greater variability in the NFR-aware code generation capabilities of LLMs.}}} Table~\ref{tab:rq1_nfr} shows that the STDEV values for \raw across the three NFR metrics (i.e., code smell, unreadability, and exception handling) are generally lower compared to the STDEV values observed in all four NFR dimensions across all models and benchmarks. For instance, when analyzing exception-handling density in HumanEval, GPT-3.5-turbo's STDEV in \raw is 0.003, which increases to 0.247 for the generated code under the Reliability dimension. Specifically, this substantial increase suggests that varying prompts significantly impact the exception-handling density of the generated code, indicating a robustness issue in LLMs when different prompts are used in the Reliability dimension. However, such a trend was not evident in the execution time metrics. For example, in the HumanEval, Claude-3.5-Haiku has a STDEV of 13.75 in the \raw, which increases to 24.43 in the Performance dimension—indicating that the generated code execution time varies more when different prompts are used to improve code performance. In contrast, Claude-3.5-Sonnet shows a STDEV of 2.93 in the \raw setting and a lower value of 1.72 in the Performance dimension, suggesting that this LLM’s generated code performance is less affected by prompt variations. This finding suggests that the impact of prompt variations on code performance differs across models, highlighting the importance of using \tool to evaluate robustness issues in different LLMs.

\rqboxc{\textbf{Summary of RQ1:} Integrating NFRs reduces Pass@1 scores by up to 39\% and increases output variability, revealing robustness issues across all models. While incorporating NFRs generally helps reduce code smells and improve readability, it also leads to higher STDEV values, indicating that code quality varies significantly with prompt variations.}

\subsection*{RQ2: Are LLMs robust to model updates in their NFR-Aware code generation capabilities?}

\label{sec:rq2}

\noindent{\underline{\textit{Motivation.}}} 
When relying on LLMs for NFR-related tasks, users often need to choose which model version to use. This is especially relevant when using the default version, which automatically switches to the latest one. In such cases, users may experience unexpected changes in outcomes due to model updates. Therefore, this study investigates how updates to LLM versions influence their ability to generate NFR-aware code.

\noindent{\underline{\textit{Approach.}}} 
We adopt Evaluation Methodology 2 (i.e., Regression Testing) as discussed in Section~\ref{sec:regression} to compare the results of functional correctness and NFR metrics among different versions of LLMs. We compute the Pass@1 and NFR metrics for four types of NFRs—code design, reliability, readability, and performance—generated by different versions of the same LLM model. We then compare the results within each model. 
Our study covers three LLM models, and for each, we compare the two most recent versions (at the time of the experiments). Specifically, we used \textit{gpt-3.5-turbo-1106} and \textit{gpt-3.5-turbo-0125} for GPT-3.5-turbo, \textit{gpt-4o-2024-05-13} and \textit{gpt-4o-2024-08-06} for GPT-4o, and \textit{claude-3-5-sonnet-20240620} and \textit{claude-3-5-haiku-20241022} for Claude-3.5. We selected these two versions from the Claude family because they share the same underlying LLM architecture and similar code generation capabilities~\cite{anthropic2024claude3}, and they were the only ones available to us.

\begin{table}[t]
\setlength{\tabcolsep}{2pt}
\centering

\caption{This table compares various metrics for the same LLM model across different versions. The Pass@1($\Delta$\%) column shows the Pass@1 score for the older version along with the percentage change relative to the newer version, while the ET-Pass@1($\Delta$\%) column presents the same metric for the ET-version dataset. Additionally, the NFR metrics—including code smell density, unreadability density, exception-handling density, and execution time (in milliseconds)—report the older version’s results with standard deviations and the corresponding percentage differences compared to the newer version.}

\scalebox{0.6}{

\begin{tabular}{lll|rr|rrrrr}\toprule
Model & Dataset & NFR Dimension & Pass@1($\Delta$\%) & ET-Pass@1($\Delta$\%) & code smell($\Delta$\%) & unreadability($\Delta$\%) & exception-handling($\Delta$\%) & execution time($\Delta$\%) \\
\midrule
\multirow{10}{*}{\shortstack{\textbf{GPT-3.5} \\ (\textit{20231106}) \\ vs \\ (\textit{20240125})}} 
  & \multirow{5}{*}{HumanEval} 
    & Function-Only 
      & 76.46±0.77 \textcolor{red}{$\downarrow$5.18} 
      & 66.83±0.51 \textcolor{red}{$\downarrow$3.74} 
      & 0.38±0.01 \textcolor{red}{$\downarrow$18.42} 
      & 2.77±0.04 \textcolor{blue}{$\uparrow$23.47} 
      & 0.011±0.003 \textcolor{blue}{$\uparrow$227.27} 
      & 110.78±46.55 \textcolor{blue}{$\uparrow$1.67} \\
  &  & Code Design 
      & 72.44±2.71 \textcolor{blue}{$\uparrow$0.59} 
      & 64.63±2.80 \textcolor{red}{$\downarrow$0.19} 
      & 0.25±0.01 \textcolor{red}{$\downarrow$12.00} 
      & 1.79±0.10 \textcolor{blue}{$\uparrow$49.72} 
      & 0.055±0.058 \textcolor{red}{$\downarrow$7.27} 
      & 97.55±49.53 \textcolor{red}{$\downarrow$4.74} \\
  &  & Readability 
      & 73.29±3.56 \textcolor{red}{$\downarrow$0.16} 
      & 64.82±2.88 \textcolor{red}{$\downarrow$0.76} 
      & 0.21±0.04 \textcolor{red}{$\downarrow$14.29} 
      & 1.58±0.09 \textcolor{blue}{$\uparrow$56.33} 
      & 0.015±0.007 \textcolor{blue}{$\uparrow$6.67} 
      & 97.58±46.15 \textcolor{blue}{$\uparrow$23.39} \\
  &  & Reliability 
      & 65.73±4.29 \textcolor{blue}{$\uparrow$3.89} 
      & 57.62±4.40 \textcolor{blue}{$\uparrow$2.55} 
      & 0.40±0.10 \textcolor{red}{$\downarrow$15.00} 
      & 1.92±0.23 \textcolor{blue}{$\uparrow$46.35} 
      & 1.362±0.311 \textcolor{red}{$\downarrow$1.47} 
      & 117.04±54.46 \textcolor{blue}{$\uparrow$0.70} \\
  &  & Performance 
      & 72.26±1.58 \textcolor{red}{$\downarrow$2.03} 
      & 63.54±2.13 \textcolor{red}{$\downarrow$2.69} 
      & 0.32±0.06 \textcolor{red}{$\downarrow$15.63} 
      & 2.41±0.18 \textcolor{blue}{$\uparrow$33.11} 
      & 0.014±0.003 \textcolor{blue}{$\uparrow$14.29} 
      & 62.87±3.01 \textcolor{blue}{$\uparrow$0.97} \\
\cmidrule(lr){2-9}
  & \multirow{5}{*}{MBPP} 
    & Function-Only 
      & 63.47±0.55 \textcolor{blue}{$\uparrow$6.85} 
      & 44.75±0.64 \textcolor{blue}{$\uparrow$5.50} 
      & 0.32±0.01 \textcolor{red}{$\downarrow$15.63} 
      & 3.64±0.02 \textcolor{red}{$\downarrow$5.49} 
      & 0.006±0.000 \textcolor{blue}{$\uparrow$83.33} 
      & 48.84±1.33 \textcolor{red}{$\downarrow$11.59} \\
  &  & Code Design 
      & 66.53±1.05 \textcolor{blue}{$\uparrow$1.73} 
      & 46.49±0.85 \textcolor{blue}{$\uparrow$2.41} 
      & 0.35±0.04 \textcolor{red}{$\downarrow$20.00} 
      & 3.71±0.18 \textcolor{blue}{$\uparrow$26.17} 
      & 0.126±0.118 \textcolor{red}{$\downarrow$7.14} 
      & 51.66±11.29 \textcolor{blue}{$\uparrow$13.78} \\
  &  & Readability 
      & 66.93±2.38 \textcolor{blue}{$\uparrow$2.73} 
      & 47.26±1.60 \textcolor{blue}{$\uparrow$2.43} 
      & 0.30±0.03 \textcolor{red}{$\downarrow$20.00} 
      & 3.42±0.28 \textcolor{blue}{$\uparrow$21.05} 
      & 0.012±0.005 \textcolor{blue}{$\uparrow$8.33} 
      & 52.02±5.54 \textcolor{red}{$\downarrow$4.94} \\
  &  & Reliability 
      & 45.11±11.71 \textcolor{red}{$\downarrow$4.83} 
      & 30.80±8.21 \textcolor{red}{$\downarrow$4.35} 
      & 0.45±0.12 \textcolor{red}{$\downarrow$20.00} 
      & 2.72±0.54 \textcolor{blue}{$\uparrow$19.49} 
      & 1.785±0.212 \textcolor{red}{$\downarrow$10.31} 
      & 42.01±3.51 \textcolor{red}{$\downarrow$2.50} \\
  &  & Performance 
      & 65.95±2.16 \textcolor{blue}{$\uparrow$1.03} 
      & 47.14±1.41 \textcolor{blue}{$\uparrow$1.44} 
      & 0.32±0.05 \textcolor{red}{$\downarrow$12.50} 
      & 5.18±0.22 \textcolor{blue}{$\uparrow$15.64} 
      & 0.011±0.003 (0.00) 
      & 47.05±6.29 \textcolor{blue}{$\uparrow$9.70} \\
\midrule
\multirow{10}{*}{\shortstack{\textbf{GPT-4o} \\ (\textit{20240513}) \\ vs \\ (\textit{20240806})}} 
  & \multirow{5}{*}{HumanEval} 
    & Function-Only 
      & 92.56±0.85 \textcolor{red}{$\downarrow$2.17} 
      & 81.52±1.00 \textcolor{red}{$\downarrow$1.64} 
      & 0.13±0.01 \textcolor{red}{$\downarrow$7.69} 
      & 2.50±0.04 \textcolor{blue}{$\uparrow$4.80} 
      & 0.040±0.002 \textcolor{red}{$\downarrow$35.00} 
      & 77.40±13.78 \textcolor{red}{$\downarrow$1.91} \\
  &  & Code Design 
      & 90.73±1.49 \textcolor{red}{$\downarrow$1.54} 
      & 80.12±1.93 \textcolor{red}{$\downarrow$0.61} 
      & 0.06±0.01 (0.00) 
      & 1.35±0.17 \textcolor{red}{$\downarrow$5.93} 
      & 0.095±0.027 \textcolor{red}{$\downarrow$4.21} 
      & 81.57±6.25 \textcolor{red}{$\downarrow$0.43} \\
  &  & Readability 
      & 92.74±1.30 \textcolor{red}{$\downarrow$1.44} 
      & 81.89±1.52 \textcolor{red}{$\downarrow$1.11} 
      & 0.07±0.01 \textcolor{red}{$\downarrow$14.29} 
      & 1.38±0.15 \textcolor{red}{$\downarrow$9.42} 
      & 0.038±0.012 \textcolor{red}{$\downarrow$23.68} 
      & 81.52±9.24 \textcolor{blue}{$\uparrow$3.64} \\
  &  & Reliability 
      & 89.09±1.92 \textcolor{red}{$\downarrow$0.90} 
      & 76.46±2.23 \textcolor{red}{$\downarrow$0.31} 
      & 0.10±0.03 (0.00) 
      & 1.66±0.19 \textcolor{red}{$\downarrow$12.65} 
      & 0.910±0.136 \textcolor{blue}{$\uparrow$3.52} 
      & 87.69±1.44 \textcolor{blue}{$\uparrow$3.18} \\
  &  & Performance 
      & 90.18±1.56 \textcolor{red}{$\downarrow$0.94} 
      & 80.73±1.63 \textcolor{red}{$\downarrow$0.68} 
      & 0.07±0.01 \textcolor{blue}{$\uparrow$14.29} 
      & 1.38±0.11 \textcolor{blue}{$\uparrow$18.84} 
      & 0.023±0.008 \textcolor{blue}{$\uparrow$17.39} 
      & 79.44±8.68 \textcolor{red}{$\downarrow$6.42} \\
\cmidrule(lr){2-9}
  & \multirow{5}{*}{MBPP} 
    & Function-Only 
      & 75.34±0.58 \textcolor{red}{$\downarrow$1.21} 
      & 53.91±0.49 \textcolor{red}{$\downarrow$1.00} 
      & 0.10±0.00 \textcolor{blue}{$\uparrow$20.00} 
      & 2.72±0.03 \textcolor{blue}{$\uparrow$16.91} 
      & 0.129±0.007 \textcolor{blue}{$\uparrow$0.78} 
      & 34.69±0.21 \textcolor{blue}{$\uparrow$7.32} \\
  &  & Code Design 
      & 73.79±0.74 \textcolor{red}{$\downarrow$0.57} 
      & 52.95±1.04 \textcolor{red}{$\downarrow$0.70} 
      & 0.06±0.01 \textcolor{red}{$\downarrow$16.67} 
      & 2.23±0.12 \textcolor{red}{$\downarrow$19.73} 
      & 0.363±0.049 (0.00) 
      & 38.64±4.47 \textcolor{red}{$\downarrow$0.88} \\
  &  & Readability 
      & 73.72±1.32 \textcolor{blue}{$\uparrow$1.79} 
      & 52.67±0.74 \textcolor{blue}{$\uparrow$1.82} 
      & 0.08±0.01 \textcolor{red}{$\downarrow$12.50} 
      & 2.26±0.10 \textcolor{red}{$\downarrow$17.70} 
      & 0.122±0.031 \textcolor{red}{$\downarrow$12.30} 
      & 37.39±2.32 \textcolor{red}{$\downarrow$2.67} \\
  &  & Reliability 
      & 71.59±0.83 (0.00) 
      & 50.35±1.01 \textcolor{red}{$\downarrow$0.66} 
      & 0.18±0.08 \textcolor{red}{$\downarrow$5.56} 
      & 2.75±0.15 \textcolor{red}{$\downarrow$5.09} 
      & 1.588±0.192 \textcolor{red}{$\downarrow$0.25} 
      & 35.44±1.81 \textcolor{red}{$\downarrow$0.99} \\
  &  & Performance 
      & 73.54±0.47 \textcolor{blue}{$\uparrow$0.72} 
      & 52.95±0.67 \textcolor{blue}{$\uparrow$1.15} 
      & 0.13±0.02 \textcolor{red}{$\downarrow$7.69} 
      & 3.31±0.19 \textcolor{red}{$\downarrow$1.51} 
      & 0.107±0.039 \textcolor{red}{$\downarrow$3.74} 
      & 36.25±2.48 \textcolor{red}{$\downarrow$4.83} \\
\midrule
\multirow{10}{*}{\shortstack{\textbf{Claude-3.5} \\ (\textit{20240620}) \\ vs \\ (\textit{20241022})}} 
  & \multirow{5}{*}{HumanEval} 
    & Function-Only 
      & 89.39±0.33 \textcolor{red}{$\downarrow$3.55} 
      & 78.54±0.51 \textcolor{red}{$\downarrow$3.73} 
      & 0.10±0.01 \textcolor{red}{$\downarrow$40.00} 
      & 2.03±0.01 \textcolor{blue}{$\uparrow$28.03} 
      & 0.037±0.003 \textcolor{red}{$\downarrow$40.54} 
      & 57.06±2.93 \textcolor{blue}{$\uparrow$23.24} \\
  &  & Code Design 
      & 84.02±1.85 \textcolor{red}{$\downarrow$3.92} 
      & 72.56±1.93 \textcolor{red}{$\downarrow$3.36} 
      & 0.02±0.01 (0.00) 
      & 0.79±0.10 \textcolor{blue}{$\uparrow$94.94} 
      & 0.148±0.064 \textcolor{blue}{$\uparrow$18.92} 
      & 47.27±15.87 \textcolor{blue}{$\uparrow$18.37} \\
  &  & Readability 
      & 86.46±1.80 \textcolor{red}{$\downarrow$2.82} 
      & 75.85±1.81 \textcolor{red}{$\downarrow$2.73} 
      & 0.04±0.02 \textcolor{red}{$\downarrow$50.00} 
      & 0.97±0.20 \textcolor{blue}{$\uparrow$42.27} 
      & 0.084±0.042 \textcolor{blue}{$\uparrow$4.76} 
      & 44.04±13.54 \textcolor{blue}{$\uparrow$39.82} \\
  &  & Reliability 
      & 88.29±1.25 \textcolor{red}{$\downarrow$17.26} 
      & 76.22±1.52 \textcolor{red}{$\downarrow$18.39} 
      & 0.05±0.01 \textcolor{red}{$\downarrow$40.00} 
      & 1.07±0.05 \textcolor{blue}{$\uparrow$64.49} 
      & 1.177±0.152 \textcolor{red}{$\downarrow$14.53} 
      & 53.22±7.62 \textcolor{blue}{$\uparrow$52.75} \\
  &  & Performance 
      & 83.29±1.53 \textcolor{red}{$\downarrow$2.37} 
      & 74.02±1.40 \textcolor{red}{$\downarrow$2.80} 
      & 0.05±0.02 \textcolor{red}{$\downarrow$60.00} 
      & 1.67±0.11 \textcolor{blue}{$\uparrow$39.52} 
      & 0.028±0.005 \textcolor{blue}{$\uparrow$17.86} 
      & 35.20±1.72 \textcolor{blue}{$\uparrow$179.50} \\
\cmidrule(lr){2-9}
  & \multirow{5}{*}{MBPP} 
    & Function-Only 
      & 75.97±0.27 \textcolor{red}{$\downarrow$4.74} 
      & 54.94±0.13 \textcolor{red}{$\downarrow$3.66} 
      & 0.08±0.00 \textcolor{red}{$\downarrow$62.50} 
      & 2.67±0.01 \textcolor{blue}{$\uparrow$0.75} 
      & 0.069±0.002 \textcolor{red}{$\downarrow$40.58} 
      & 40.92±3.68 \textcolor{red}{$\downarrow$15.93} \\
  &  & Code Design 
      & 70.87±2.58 \textcolor{red}{$\downarrow$8.66} 
      & 49.79±2.23 \textcolor{red}{$\downarrow$8.27} 
      & 0.03±0.01 \textcolor{red}{$\downarrow$66.67} 
      & 1.66±0.18 \textcolor{blue}{$\uparrow$17.47} 
      & 0.454±0.142 \textcolor{red}{$\downarrow$1.98} 
      & 36.42±0.18 \textcolor{red}{$\downarrow$3.40} \\
  &  & Readability 
      & 73.35±2.64 \textcolor{red}{$\downarrow$16.09} 
      & 51.66±1.52 \textcolor{red}{$\downarrow$14.23} 
      & 0.05±0.02 \textcolor{red}{$\downarrow$60.00} 
      & 1.54±0.24 \textcolor{blue}{$\uparrow$4.55} 
      & 0.261±0.097 \textcolor{red}{$\downarrow$13.41} 
      & 39.04±8.94 \textcolor{red}{$\downarrow$9.50} \\
  &  & Reliability 
      & 71.59±0.88 \textcolor{red}{$\downarrow$33.72} 
      & 50.02±0.70 \textcolor{red}{$\downarrow$35.95} 
      & 0.05±0.01 \textcolor{red}{$\downarrow$80.00} 
      & 1.98±0.45 \textcolor{red}{$\downarrow$25.25} 
      & 1.354±0.107 \textcolor{red}{$\downarrow$17.64} 
      & 43.66±15.62 \textcolor{red}{$\downarrow$20.54} \\
  &  & Performance 
      & 72.04±1.76 \textcolor{red}{$\downarrow$1.82} 
      & 51.43±2.08 \textcolor{red}{$\downarrow$3.93} 
      & 0.05±0.00 \textcolor{red}{$\downarrow$60.00} 
      & 2.51±0.12 \textcolor{red}{$\downarrow$3.59} 
      & 0.096±0.047 \textcolor{red}{$\downarrow$27.08} 
      & 34.62±0.64 \textcolor{blue}{$\uparrow$7.42} \\
\bottomrule
\end{tabular}
}
\label{tab:rq2}
\vspace{-3mm}
\end{table}

\noindent{\underline{\textit{Results.}}} 
\textit{\textbf{When model version updates occur, trade-offs emerge between code correctness and NFR metrics.}} As illustrated in Table~\ref{tab:rq2}, improvements in one area—such as a reduction in code smell density—are often offset by declines in other metrics, like Pass@1 or ET-Pass@1. For instance, in GPT-3.5-turbo on HumanEval under the \raw, the pass rate drops by 5.18\% and 3.74\%, while the code smell metric improves significantly by 18.42\%. This pattern suggests that tuning to enhance certain qualitative aspects may come at the expense of Pass@1, and vice versa.

Interestingly, not every model clearly exhibits trade-off patterns within the same metric groups. \textit{\textbf{Some models appear to be less robust, such as GPT-3.5-turbo, which exhibits inconsistent trade-off patterns across benchmarks, indicating greater robustness issues in these LLMs.}} As shown in Table~\ref{tab:rq2}, the impact of certain model updates is clearly not uniform across NFR dimensions or datasets. For instance, the metric groups Pass@1 and unreadability for newer GPT-3.5-turbo show inconsistent trade-off patterns. Specifically, on the MBPP dataset, the newer GPT-3.5-turbo shows an improvement in \raw, with a 6.85\% increase in Pass@1 scores and a 5.49\% decrease in unreadability density, suggesting improved correctness and code quality compared to its older version. In contrast, on the HumanEval dataset, the newer GPT-3.5-turbo shows a decline in \raw, with a 5.18\% decrease in Pass@1 and a 23.47\% increase in unreadability density, indicating that it performs worse in both aspects compared to the older one. This inconsistent trend across datasets highlights potential robustness issues in GPT-3.5-turbo updates, making it more difficult to accurately assess the LLM's NFR-aware code generation capabilities. Notably, some models, like GPT-4o, appear to be more robust, as their trade-off patterns are more consistent. For the same metric group—Pass@1 and unreadability—the newer GPT-4o consistently achieves lower Pass@1 scores and higher unreadability density, as shown in Table~\ref{tab:rq2}. Although these results indicate that the newer GPT-4o performs worse than its older version, they clearly highlight how its NFR-aware code generation abilities change with model updates. This insight can help users make informed decisions about its usage and assist LLM developers in debugging.

\textbf{\textit{Additionally, GPT-4o exhibits smaller percentage changes overall, indicating more stable NFR-aware code generation capabilities when updated to a new LLM version.}} For example, when examining Pass@1 for GPT-4o across all NFR dimensions, the percentage change ranges from 0.00\% to 2.17\%, compared to GPT-3.5, which ranges from 0.59\% to 6.85\%, and Claude-3.5, which ranges from 1.82\% to 33.72\%. This indicates that when GPT-4o is updated, it experiences only minor variations on both the HumanEval and MBPP datasets. Compared to GPT-3.5 and Claude-3.5, the results suggests that GPT-4o is more robust to version updates among the models we tested.

As discussed above, GPT-4o appears to be the most robust among the three models, which is evident when examining only the results from the Reliability dimension. This means that \textit{\textbf{the Reliability dimension is the most effective in identifying robustness issues in LLMs' NFR-aware code generation capabilities when the model is updated.}} As shown in Table~\ref{tab:rq2}, Claude-3.5 exhibited significant changes, with a 17.26\% decrease in Pass@1 on HumanEval and a 33.72\% decrease on MBPP, while GPT-3.5 also showed percentage changes of 3.89\% on HumanEval and 4.83\% on MBPP. In contrast, GPT-4o displayed minimal changes, with only a 0.90\% change on HumanEval and no change (0.00\%) on MBPP. These smaller percentage changes indicate that when the model is updated, GPT-4o is more robust compared to the other two models, aligning with the findings discussed above. A potential reason is that the Reliability dimension requires LLMs to add exception handling, which typically alters the code structure more and introduces additional statements compared to other NFR dimensions. 

Overall, these findings highlight that different LLM models exhibit varying changes in NFR-aware code generation capabilities when updated. This underscores the importance of using robustness evaluation frameworks like \tool to identify potential issues in LLMs, whether for user adoption or developer debugging.

\rqboxc{\textbf{Summary of RQ2:} When LLMs are updated, they introduce trade-offs between code correctness and NFR metrics. Some models are less robust—such as GPT-3.5-turbo exhibits inconsistent trade-offs after an update—while GPT-4o appears more robust, as it maintains more consistent trade-offs and exhibits smaller percentage changes across all metrics. Among the NFR dimensions, the Reliability dimension serves as a strong indicator for monitoring LLM robustness in NFR-aware code generation, as it represents the degree of change between newer and older versions of the same model.}

\subsection*{RQ3: Do robustness issues exist in LLMs' NFR-aware code generation when different workflows are applied?}

\noindent{\underline{\textit{Motivation.}}} 
As discussed in Section~\ref{sec:approach}, users primarily rely on \first or \second when using LLMs for NFR dimensions. Therefore, this research question investigates the NFR-aware code generation capabilities of the two workflows, comparing them based on both code correctness (measured by Pass@1) and NFRs metrics.

\noindent{\underline{\textit{Approach.}}} 
We adopt Evaluation Methodology 3 (i.e., NFR-Aware Code Generation Workflows) as discussed in Section~\ref{sec:approach}. For this experiment, we use the same models as in RQ2 (see Section~\ref{sec:rq2}), namely \gptThreeOld, \gptThreeNew, \gptFourOld, \gptFourNew, \claudeOld, and \claudeNew. We also use the 10 prompts discussed in RQ1 (see Section~\ref{sec:rq1}) to compute the average Pass@1 and the standard deviation (STDEV). Finally, we average these values to obtain the final comparison.

\begin{table}[t]
\setlength{\tabcolsep}{2pt}
\caption{Pass@1 AVG: This column shows the average Pass@1 score computed across all experimental models. $\Delta$: The $\Delta$ symbol indicates the Pass@1 difference between the results of various NFR dimensions and the baseline (\raw) results.}

\centering
\scalebox{0.82}{
\begin{tabular}{ll|rr|rr|rr|rr}\toprule
\multirow{2}{*}{Workflow} &\multirow{2}{*}{NFR Dimension} &\multicolumn{2}{c}{HumanEval} &\multicolumn{2}{c}{HumanEval-ET} &\multicolumn{2}{c}{MBPP} &\multicolumn{2}{c}{MBPP-ET} \\\cmidrule{3-10}
& &Pass@1 AVG &$\Delta$(\%) &Pass@1 AVG &$\Delta$(\%) &Pass@1 AVG &$\Delta$(\%) &Pass@1 AVG &$\Delta$(\%) \\\midrule
- & \raw &84.61±0.70 &0 &74.50±0.74 &0 &71.57±0.41 &0 &51.19±0.33 &0 \\
\midrule

\multirow{4}{*}{NFR-Integrated} 
    & Code Design & 81.69±1.84 & \textcolor{red}{$\downarrow$3.46} & 71.93±2.09 & \textcolor{red}{$\downarrow$3.45} & 69.50±2.06 & \textcolor{red}{$\downarrow$2.89} & 49.18±1.71 & \textcolor{red}{$\downarrow$3.91} \\
    & Readability & 83.51±2.42 & \textcolor{red}{$\downarrow$1.30} & 73.61±2.16 & \textcolor{red}{$\downarrow$1.20} & 69.89±2.75 & \textcolor{red}{$\downarrow$2.34} & 49.66±1.89 & \textcolor{red}{$\downarrow$2.99} \\
    & Reliability & 77.71±2.64 & \textcolor{red}{$\downarrow$8.15} & 66.95±2.79 & \textcolor{red}{$\downarrow$10.13} & 58.00±5.56 & \textcolor{red}{$\downarrow$18.96} & 39.91±4.03 & \textcolor{red}{$\downarrow$22.03} \\
    & Performance & 81.32±1.98 & \textcolor{red}{$\downarrow$3.89} & 72.04±1.78 & \textcolor{red}{$\downarrow$3.30}  & 70.49±1.57 & \textcolor{red}{$\downarrow$1.50}  & 50.39±1.47 & \textcolor{red}{$\downarrow$1.56} \\
\midrule

\multirow{4}{*}{NFR-Enhanced} 
    & Code Design & 72.43±7.54 & \textcolor{red}{$\downarrow$14.40} & 64.02±6.84 & \textcolor{red}{$\downarrow$14.07} & 53.48±12.12 & \textcolor{red}{$\downarrow$25.27} & 37.73±8.64 & \textcolor{red}{$\downarrow$26.28} \\
    & Readability & 76.05±5.96 & \textcolor{red}{$\downarrow$10.12} & 66.87±5.29 & \textcolor{red}{$\downarrow$10.25} & 57.41±8.08 & \textcolor{red}{$\downarrow$19.78} & 40.67±5.51 & \textcolor{red}{$\downarrow$20.55} \\
    & Reliability & 72.75±4.54 & \textcolor{red}{$\downarrow$14.02} & 62.09±4.36 & \textcolor{red}{$\downarrow$16.66} & 61.04±4.08 & \textcolor{red}{$\downarrow$14.71} & 42.38±2.93 & \textcolor{red}{$\downarrow$17.21} \\
    & Performance & 77.02±2.70 & \textcolor{red}{$\downarrow$8.97} & 68.23±2.49 & \textcolor{red}{$\downarrow$8.41} & 68.53±2.01 & \textcolor{red}{$\downarrow$4.24} & 48.98±1.67 & \textcolor{red}{$\downarrow$4.30} \\
\bottomrule
\end{tabular}
}
\label{tab:rq3_pass@1}
\vspace{-3mm}
\end{table}

\noindent{\underline{\textit{Results.}}} 
\begin{table}[t]

\caption{Columns \textit{code smell density}, \textit{unreadability density}, \textit{exception-handling density}, and \textit{execution time (millisecond)} represent the NFR metrics (Section~\ref{metrics}). Each metric includes standard deviations and $\Delta$\%, which indicates the percentage difference between NFR-aware results and \raw results.}

\setlength{\tabcolsep}{1mm}
\centering
\scalebox{1}{
\scriptsize
\begin{tabular}{lcl|rrrr}\toprule
Dataset &Workflow & NFR Dimension &code smell($\Delta$\%) &unreadability($\Delta$\%) &exception-handling($\Delta$\%) &execution time($\Delta$\%) \\\midrule
\multirow{9}{*}{HumanEval} &- &Function-Only &0.18±0.01 &2.66±0.03 &0.029±0.003 &84.02±21.55 \\
\cmidrule(lr){2-7}
&\multirow{4}{*}{NFR-Integrated} &Code Design &0.10±0.01 (\(\downarrow \textcolor{red}{44.4}\)) &1.57±0.20 (\(\downarrow \textcolor{red}{41.0}\)) &0.103±0.054 (\(\uparrow \textcolor{blue}{255.2}\)) &76.08±18.54 (\(\downarrow \textcolor{red}{9.45}\)) \\
& &Readability &0.10±0.02 (\(\downarrow \textcolor{red}{44.4}\)) &1.51±0.16 (\(\downarrow \textcolor{red}{43.2}\)) &0.045±0.019 (\(\uparrow \textcolor{blue}{55.2}\)) &81.60±22.35 (\(\downarrow \textcolor{red}{2.88}\)) \\
& &Reliability &0.17±0.05 (\(\downarrow \textcolor{red}{5.6}\)) &1.78±0.20 (\(\downarrow \textcolor{red}{33.1}\)) &1.123±0.180 (\(\uparrow \textcolor{blue}{3772.4}\)) &91.27±25.12 (\(\uparrow \textcolor{blue}{8.63}\)) \\
& &Performance &0.14±0.02 (\(\downarrow \textcolor{red}{22.2}\)) &2.11±0.14 (\(\downarrow \textcolor{red}{20.7}\)) &0.024±0.006 (\(\downarrow \textcolor{red}{17.2}\)) &68.95±12.11 (\(\downarrow \textcolor{red}{17.94}\)) \\
\cmidrule(lr){2-7}
&\multirow{4}{*}{NFR-Enhanced} &Code Design &0.06±0.02 (\(\downarrow \textcolor{red}{66.7}\)) &1.23±0.20 (\(\downarrow \textcolor{red}{53.8}\)) &0.090±0.039 (\(\uparrow \textcolor{blue}{210.3}\)) &73.03±26.91 (\(\downarrow \textcolor{red}{13.08}\)) \\
& &Readability &0.07±0.02 (\(\downarrow \textcolor{red}{61.1}\)) &1.24±0.15 (\(\downarrow \textcolor{red}{53.4}\)) &0.056±0.013 (\(\uparrow \textcolor{blue}{93.1}\)) &75.14±24.52 (\(\downarrow \textcolor{red}{10.57}\)) \\
& &Reliability &0.09±0.02 (\(\downarrow \textcolor{red}{50.0}\)) &1.41±0.16 (\(\downarrow \textcolor{red}{47.0}\)) &0.855±0.150 (\(\uparrow \textcolor{blue}{2848.3}\)) &81.50±27.27 (\(\downarrow \textcolor{red}{3.00}\)) \\
& &Performance &0.10±0.02 (\(\downarrow \textcolor{red}{44.4}\)) &1.62±0.10 (\(\downarrow \textcolor{red}{39.1}\)) &0.035±0.011 (\(\uparrow \textcolor{blue}{20.7}\)) &78.41±18.92 (\(\downarrow \textcolor{red}{6.68}\)) \\[1ex]
\midrule
\multirow{9}{*}{MBPP} &- &Function-Only &0.15±0.01 &3.06±0.02 &0.064±0.002 &39.88±2.47 \\
\cmidrule(lr){2-7}
&\multirow{4}{*}{NFR-Integrated} &Code Design &0.13±0.02 (\(\downarrow \textcolor{red}{13.3}\)) &2.67±0.28 (\(\downarrow \textcolor{red}{12.7}\)) &0.311±0.103 (\(\uparrow \textcolor{blue}{385.9}\)) &43.16±5.32 (\(\uparrow \textcolor{blue}{8.22}\)) \\
& &Readability &0.12±0.02 (\(\downarrow \textcolor{red}{20.0}\)) &2.47±0.23 (\(\downarrow \textcolor{red}{19.3}\)) &0.124±0.043 (\(\uparrow \textcolor{blue}{93.8}\)) &41.60±5.15 (\(\uparrow \textcolor{blue}{4.31}\)) \\
& &Reliability &0.20±0.06 (\(\uparrow \textcolor{blue}{33.3}\)) &2.47±0.36 (\(\downarrow \textcolor{red}{19.3}\)) &1.504±0.169 (\(\uparrow \textcolor{blue}{2250.0}\)) &38.64±3.72 (\(\downarrow \textcolor{red}{3.11}\)) \\
& &Performance &0.15±0.02 (\(\downarrow \textcolor{red}{0.0}\)) &3.78±0.19 (\(\uparrow \textcolor{blue}{23.5}\)) &0.066±0.024 (\(\uparrow \textcolor{blue}{3.1}\)) &40.21±3.88 (\(\uparrow \textcolor{blue}{0.83}\)) \\
\cmidrule(lr){2-7}
&\multirow{4}{*}{NFR-Enhanced} &Code Design &0.05±0.02 (\(\downarrow \textcolor{red}{66.7}\)) &2.18±0.46 (\(\downarrow \textcolor{red}{28.8}\)) &0.224±0.059 (\(\uparrow \textcolor{blue}{250.0}\)) &43.02±5.56 (\(\uparrow \textcolor{blue}{7.87}\)) \\
& &Readability &0.07±0.02 (\(\downarrow \textcolor{red}{53.3}\)) &2.36±0.23 (\(\downarrow \textcolor{red}{22.9}\)) &0.152±0.031 (\(\uparrow \textcolor{blue}{137.5}\)) &44.15±6.64 (\(\uparrow \textcolor{blue}{10.71}\)) \\
& &Reliability &0.13±0.05 (\(\downarrow \textcolor{red}{13.3}\)) &2.23±0.24 (\(\downarrow \textcolor{red}{27.1}\)) &1.327±0.182 (\(\uparrow \textcolor{blue}{1973.4}\)) &41.12±3.79 (\(\uparrow \textcolor{blue}{3.11}\)) \\
& &Performance &0.13±0.02 (\(\downarrow \textcolor{red}{13.3}\)) &3.14±0.19 (\(\uparrow \textcolor{blue}{2.6}\)) &0.089±0.036 (\(\uparrow \textcolor{blue}{39.1}\)) &41.25±2.32 (\(\uparrow \textcolor{blue}{3.44}\)) \\
\bottomrule
\end{tabular}

}
\label{tab:rq3_nfr}
\vspace{-5mm}
\end{table}
{\textit{\textbf{\first almost always achieves better Pass@1 than \second.}}} In Table \ref{tab:rq3_pass@1}, our finding shows that a two-step approach has a negative impact on Pass@1, and the difference can be over 20\% (e.g., between \first and \second for Code Design in MBPP), depending on the specific NFR and dataset. 
For code design and readability,  the decrease is even more notable in \second (10\% to over 20\% compared to \raw) compared to \first (1.3\% to 3.91\% over \raw). In contrast, even though exception handling (i.e., Reliability) has the largest decrease in \first, the difference with \second is smaller. Performance has relatively more stable results between \first and \second. As \first shows an average 2.56\% decrease compared to \raw in the performance dimension, and \second shows a 6.48\% decrease, the gap between them is only 3.92\%—much smaller than in other NFR dimensions (e.g., Code Design shows a 16.58\% gap between \first and \second). 

Our findings show that the one-step approach may allow the LLM to balance the objectives better, and \textit{\textbf{generative models may perform worse at Pass@1 on a two-step code enhancement, especially if the NFR is more related to re-structuring the code (i.e., code design and readability)}}. 

\textit{\textbf{Incorporating NFRs reduces the capability of LLMs in stably generating functionally correct code, resulting in more variable Pass@1.}} 
\first and \second consistently exhibit higher standard deviations (STDEV) of Pass@1 across all benchmarks compared to \raw. For example, in HumanEval, the STDEV for Pass@1 ranges from 1.84 to 2.64 for \first and 2.70 to 7.54 for \second, both much higher than the STDEV of 0.70 for \raw. Moreover, we find that \second exhibits higher variability in Pass@1 than \first, \textit{\textbf{which aligns with our earlier finding that LLMs are better at generating functionally correct code in one-step approach}}. 

\textit{\textbf{Unlike Pass@1, \second leads to a larger improvement in certain non-functional code quality than \first.}} While \first outperforms \second in Pass@1, \second excels in improving NFR metrics. For code smell density, \first achieves a reduction of 13.3\% and 44.4\% on HumanEval and MBPP, respectively, whereas \second reduces by 66.7\% for both datasets. Similarly, for readability, \first improves by 19.3\%--43.2\%, while \second achieves 22.9\%--53.4\% enhancements. 
Interestingly, an inverse pattern emerges for reliability, where \first outperforms \second with improvements of 2250.0\%–3772.4\% for HumanEval and MBPP, compared to \second's 1973.4\%–2848.3\%. A similar trend is observed for the performance metric, with \first reducing execution time by 17.94\% compared to \second's 6.68\% in HumanEval, but no statistically significant difference in MBPP (t-test's p-value > 0.05).  

Our findings suggest that the two NFR-aware code generation workflows have varying benefits depending on the NFRs. While \second is more effective for improving readability and reducing code designs, \first may be better suited for addressing runtime-related requirements like exception handling and performance.

\textbf{\textit{On average, \first and \second share similar levels of stability in the NFR metrics.}}  
\raw has the lowest STDEV across all NFR metrics, partly because of its lack of consideration of NFRs. In comparison, \first and \second have larger STDEVs, but the values are often stable. For example, code smell density has an STDEV of 0.01--0.02, and unreadability density has an STDEV of 0.15--0.23 for both NFR-aware workflows. 

Overall, as revealed by \tool, even minor differences in workflows—such as executing a process like \first versus \textit{NFR-Enhanced}—can lead to different outcomes. This highlights the importance of future research in designing clear workflows for deploying LLMs and evaluating their capabilities in NFR-aware code generation.

% Future research should consider regression testing and data selection strategies during fine-tuning and model training to consider NFR and improve stability. 

\rqboxc{\textbf{Summary of RQ3:} When incorporating NFRs, \first consistently generates functionally correct code more frequently than \second. While both approaches improve the relevant metrics, \second excels in readability and code structure, whereas \first demonstrates superior performance in exception handling and runtime efficiency.}

\section{Discussion}\label{sec:discussion}

%Moreover, variability in the capability across different LLM versions highlights a critical sensitivity to workflow design and prompt construction. Earlier model versions exhibit greater variability in both functional and non-functional metrics, underscoring the importance of robust regression testing as LLMs continue to evolve.

\subsection{Discussion of Implications}

Our findings highlight implications for two key groups of stakeholders: (i) practitioners and (ii) LLM researchers.

\noindent\textbf{For Practitioners.} 
Our experiments with \tool revealed various robustness issues in LLMs that practitioners must consider, such as the fact that different prompts expressing the same meaning can significantly affect NFR-aware code generation capability, making it crucial to carefully experiment with and select the most effective prompt. Additionally, since the NFR-aware code generation ability of LLMs may vary across versions, users should either commit to a specific model version instead of automatically adopting the latest release, or reevaluate their choice whenever an update occurs. Moreover, even minor differences in workflows—such as executing a process in a single iteration (i.e., \first) versus sequentially (i.e., \second)—can lead to different outcomes. Finally, our comparisons of functional correctness and NFR metrics demonstrate that balancing competing objectives (e.g., Pass@1 versus non-functional code quality) is essential. 

In practice, to ensure the expected outcomes of LLM-based products, it is important to establish a continuous quality assurance mechanism—for example, using \tool to monitor the robustness of deployed LLMs in real-world scenarios and prevent unexpected changes in product behavior.

\noindent\textbf{For LLM Researchers.} 
The observed robustness issues and trade-offs between functional correctness and non-functional quality point to key directions for improving research setups and training processes. Future LLM studies should report the prompts used, the specific model version (including its release date), and detailed workflow information, as our experiments demonstrate that these factors significantly impact the NFR-aware code generation capabilities of LLMs. Moreover, research should focus on developing models that effectively address both functional and non-functional requirements to reduce trade-offs and variability in functional and non-functional code quality. Future studies should also monitor robustness across various code quality aspects and benchmarks to optimize training and better align with complex software requirements.

\subsection{Discussion of Failure Examples When The LLM Attempts To Address Both Functional And
Non-Functional Requirements}

To provide more insights on the robustness issues, we present some code examples exposed by \tool, demonstrating that LLMs make some mistakes when addressing NFRs, such as reliability, readability, performance, and code design.

\subsubsection{Reliability}

To enhance reliability, LLMs often include additional exception-handling statements in the code. However, they sometimes make errors, such as using incorrect try-except formats or raising generic exceptions instead of returning specific results as described in the problem requirements. 

For example, in Code~\ref{lst:HumanEval-157}, the LLM introduced extra exception-handling code that conflicts with the specified functional requirements. The error-handling code introduces extra checks for the triangle inequality theorem on lines 23-24, leading to failure. Specifically, the function's docstring states that such inputs should return False. However, the error-handling logic raises a ValueError instead, resulting in failure. This discrepancy highlights that while NFR-aware code generation can produce functionally correct logic, it may conflict with the specified evaluation criteria. Furthermore, such unexpected outputs might unfairly lead to a significant decrease in Pass@1 scores during LLM evaluation, suggesting a potential direction for designing fairer benchmarks in the future.

\begin{lstlisting}[language=PYTHON,caption={HumanEval-157 Test Failure Due to \textit{raise ValueError} Which Violates the Functional Requirement.},label={lst:HumanEval-157}, basicstyle=\scriptsize\ttfamily]
def right_angle_triangle(a, b, c):
    '''
    Given the lengths of the three sides of a triangle. Return True if the three
    sides form a right-angled triangle, False otherwise.
    A right-angled triangle is a triangle in which one angle is right angle or 
    90 degree.

    Example:
    right_angle_triangle(3, 4, 5) == True
    right_angle_triangle(1, 2, 3) == False
    '''
    
    # Type checking
    for side in (a, b, c):
        if not isinstance(side, (int, float)):
            raise TypeError(f"Invalid type: {side}. All sides must be numbers (int or float).")
    
    # Value checking
    if a <= 0 or b <= 0 or c <= 0:
        raise ValueError("All sides must be positive numbers.")
    
    # Triangle inequality theorem
    if a + b <= c or a + c <= b or b + c <= a:
        raise ValueError("The given sides do not form a triangle.") 
    
    # Check for right angle using Pythagorean theorem
    sides = sorted([a, b, c])  # Sort the sides to identify the longest side
    return sides[0]**2 + sides[1]**2 == sides[2]**2
\end{lstlisting}

\subsubsection{Code Design}
When provided with only functional requirements, the LLM can generate correct code; however, the generated code may still exhibit code smell issues. For instance, as shown in Code~\ref{lst:HumanEval-124-1}, the function-only solution passes the evaluation but relies heavily on multiple if-else statements for validation checks, introducing potential maintainability concerns.

\begin{lstlisting}[language=PYTHON,caption={HumanEval-124 correct but has code smell.},label={lst:HumanEval-124-1}, basicstyle=\scriptsize\ttfamily]
def valid_date(date):
    """You have to write a function which validates a given date string and
    returns True if the date is valid otherwise False.
    The date is valid if all of the following rules are satisfied:
    1. The date string is not empty.
    2. The number of days is not less than 1 or higher than 31 days for months 1,3,5,7,8,10,12. And the number of days is not less than 1 or higher than 30 days for months 4,6,9,11. And, the number of days is not less than 1 or higher than 29 for the month 2.
    3. The months should not be less than 1 or higher than 12.
    4. The date should be in the format: mm-dd-yyyy
    for example: 
    valid_date('03-11-2000') => True
    valid_date('15-01-2012') => False
    valid_date('04-0-2040') => False
    valid_date('06-04-2020') => True
    valid_date('06/04/2020') => False
    """
    # Check if the date string is empty
    if not date:
        return False
    # Check if the date string is in the correct format
    if len(date) != 10 or date[2] != '-' or date[5] != '-':
        return False
    # Split the date string into month, day, and year
    try:
        month, day, year = map(int, date.split('-'))
    except ValueError:
        return False
    # Check if the month is valid
    if month < 1 or month > 12:
        return False
    # Check if the day is valid based on the month
    if month in {1, 3, 5, 7, 8, 10, 12}:
        if day < 1 or day > 31:
            return False
    elif month in {4, 6, 9, 11}:
        if day < 1 or day > 30:
            return False
    elif month == 2:
        if day < 1 or day > 29:
            return False
    # If all checks pass, the date is valid
    return True
\end{lstlisting}

When processing NFR-aware code generation, LLMs may have the capability to avoid code smells, but they can sometimes make mistakes. For instance, as shown in Code~\ref{lst:HumanEval-124}, when prompted to avoid a specific code smell, the LLM attempted to mitigate it by using a regular expression pattern. However, it failed to produce a correct and functional pattern. These inaccuracies resulted in AssertionError exceptions during evaluation when the function was executed. This highlights the potential risk that an LLM might alter the original intent when generating regular expressions. Future research could explore how robustly LLMs can handle regular expression logic during code generation.

\begin{lstlisting}[language=PYTHON,caption={HumanEval-124 Test Failure while LLM using re to avoid code smell but introduce other error.},label={lst:HumanEval-124}, basicstyle=\scriptsize\ttfamily]
import re
def valid_date(date):
    # Check if the date string is empty
    if not date:
        return False
        
    # Define the regex pattern for the date format
    date_pattern = re.compile(r'^(0[1-9]|1[0-2])-(0[1-9]|[12][0-9]|3[01])-(\d{4})$')
    
    # Check if the date matches the pattern
    match = date_pattern.match(date)
    if not match:
        return False
        
    # Extract month, day, and year from the date
    month, day, year = map(int, match.groups())
    
    # Define the number of days in each month
    days_in_month = {
        1: 31, 2: 29, 3: 31, 4: 30, 5: 31, 6: 30,
        7: 31, 8: 31, 9: 30, 10: 31, 11: 30, 12: 31
    }
    
    # Check if the day is valid for the given month
    if day < 1 or day > days_in_month[month]:
        return False
    return True
assert candidate("03-15-957") == True # match fail 
\end{lstlisting}

\subsubsection{Readability}

Similar to its ability to address code design, LLMs can enhance the readability of generated code. However, we observed that LLMs sometimes prioritize suggesting strategies over producing more readable code. For example, as shown in Code~\ref{lst:HumanEval-40}, while the LLM attempted to improve code readability, it occasionally failed to generate valid code. Instead, it returned strategy descriptions as responses, which led to evaluation failures. When executed, this resulted in a SyntaxError during evaluation. During our evaluation, we observed that earlier models (e.g., \textit{GPT-3.5-turbo}) more frequently generated instructional responses instead of actual code, as illustrated here. This suggests that, to ensure the robustness of LLMs in code generation tasks, the development process should include test cases that verify whether the model’s output contains valid code.

\begin{lstlisting}[language=PYTHON,caption={HumanEval-40 Test Failure while LLM improve readability.},label={lst:HumanEval-40}, basicstyle=\scriptsize\ttfamily]
One way to enhance readability and improve the code is to add comments to explain the logic and steps of the algorithm. This will make it easier for others (and yourself) to understand the code.
Another way to improve the code is to use more descriptive variable names. For example, instead of using "l" as the variable name for the list, you could use a more descriptive name such as "input_list".
Additionally, you could extract the logic inside the function into smaller, more descriptive functions. For example, you could create a function to check if a triple of elements sum to zero, and then call that function inside the main function.
Lastly, you could add type hints to the function signature to make it clear what type of input the function expects and what type of output it returns. This can improve readability and help catch potential errors early on.
\end{lstlisting}

\subsubsection{Performance}
In the absence of specific requirements for code generation, LLMs may prioritize generating correct code while overlooking other critical factors, such as execution time. For instance, as shown in Code~\ref{lst:HumanEval-41-1}, the \raw solution implements an algorithm with $O(n^2)$ time complexity. While the code is functionally correct, it is computationally inefficient.

\begin{lstlisting}[language=PYTHON,caption={HumanEval-41, Function-only-generated code, correct but slow.},label={lst:HumanEval-41-1}, basicstyle=\scriptsize\ttfamily]
def car_race_collision(n: int):
    """
    Simulates the car collisions explicitly and calculates the number of collisions.
    Arguments:
    n -- Number of cars in each group (left-to-right and right-to-left)
    Returns:
    int -- Number of collisions
    """
    
    # Initialize collision count
    collisions = 0
    
    # Iterate through each car in the left-to-right group
    for _ in range(n):
        # Each car collides with all cars in the right-to-left group
        for _ in range(n):
            collisions += 1
    return collisions
\end{lstlisting}

When tasked with optimizing an algorithm's runtime, the LLM occasionally makes mistakes, such as using incorrect formulas to solve the problem. For instance, in Code~\ref{lst:HumanEval-41-2}, the correct formula for calculating the number of collisions is $n \times n$. However, the LLM sometimes employed alternative mathematical formulas, resulting in incorrect outputs. During evaluation, these inaccuracies led to AssertionError exceptions when the function was executed. Research on using LLMs to enhance the execution speed of existing code may benefit from adding checkpoints to address such issues—for example, by prompting the LLM to verify whether the generated mathematical formulas align with the original functional requirements.

\begin{lstlisting}[language=PYTHON,caption={HumanEval-41 Test Failure while LLM improve performance but use wrong formula.},label={lst:HumanEval-41-2}, basicstyle=\scriptsize\ttfamily]
# Correct-Code:
def car_race_collision(n: int):
    return n * n

    
# LLM NFR-aware Generated Code: Efficient but Incorrect
def car_race_collision(n: int):
    return n * (n - 1) // 2
\end{lstlisting}

\section{Threats to Validity}\label{sec:threats}

\phead{Internal Threats.}
%\subsection{Limited evaluation focus}
The primary objective of our framework is to assess LLM robustness in NFR-aware code generation by evaluating both Pass@1 and non-functional code quality. Although \tool is not explicitly pre-trained for code refinement, it mirrors how developers use LLMs—including prompt design, model version, and workflow choices—for both code generation and refinement tasks. The insights from our evaluation can inform improvements in future model architectures, guide the prioritization of code optimization efforts, and help develop strategies for more effective and robust handling of non-functional requirements in generated code. Future work could explore refining training processes or implementing targeted NFR optimization techniques during code generation and refinement.

\phead{External Threats.}
%\subsection{Limited set of LLMs}
We use a certain set of widely used LLMs to conduct the experiments. The results may not apply to all models, as results may vary across different architectures and training methods. Future studies could benefit from incorporating a broader range of models to validate the results.
%\subsection{Limited evaluation of programming languages}
In this study, we have primarily examined Python datasets. While Python is a widely used language, the generalizability of our framework to other programming languages remains to be fully explored. However, our framework is not inherently language-specific. It is expected to be applicable to other languages and can be further verified by future studies.

\section{Conclusion}\label{sec:conclusion}
This study examines the challenges and opportunities of integrating non-functional requirements (NFRs) into code generation using large language models (LLMs). We introduce \tool, a generalizable framework for evaluating LLM robustness in NFR-aware code generation by incorporating prompt variations, regression testing, and diverse workflows for leveraging LLM capabilities. Our findings reveal potential robustness issues and significant trade-offs between functional correctness and non-functional code quality attributes such as design, readability, reliability, and performance.

Our study demonstrates that while incorporating NFRs into code generation reduces the functional correctness metric (e.g., Pass@1), it yields notable improvements in non-functional code quality. However, we also observed a heightened potential for robustness issues. Our analysis of three evaluation methodologies highlights the importance of robustness: The selection of semantically equivalent prompts can significantly impact both functional and NFR metrics; model updates require regression testing to maintain consistency; and different workflows—such as \first and \second—demonstrate varying degrees of effectiveness in addressing specific NFR aspects. By providing real-time feedback, \tool facilitates continuous monitoring of LLM robustness and helps identify the optimal combination of prompt design, model version, and workflow selection, ultimately enhancing LLM-based solutions and development efficiency.

\end{document}